\pdfoutput=1
\documentclass[
nofootinbib,
,amsfonts,amssymb,amsmath, twocolumn
]{revtex4-2}
\usepackage[utf8]{inputenc}
\usepackage[T1]{fontenc}
\usepackage[oldstyle]{libertine}
\usepackage{libertinust1math} 
\usepackage[english]{babel}
\usepackage{fontawesome}
\usepackage{amsmath,amssymb,amsfonts,bbold,bbm,bm,graphicx,mathtools,perpage,physics} 
\usepackage[scr=boondoxo,bb=boondox]{mathalfa} 
\usepackage[margin=10pt,font=small,labelfont=bf]{caption}
\usepackage[pdftex,dvipsnames]{xcolor}
\definecolor{darkblue}{rgb}{0.1,0.1,.7}
\usepackage[colorlinks, linkcolor=darkblue, citecolor=darkblue, urlcolor=darkblue, linktocpage]{hyperref} 
\numberwithin{equation}{section}

\graphicspath{ {/} }
\newcommand{\equref}[1]{eqn.~\eqref{#1}}

\def\<#1\>{\expval{#1}}

\newcommand   \tl  [1] {\widetilde{#1}}
\renewcommand \.   {\cdot}
\newcommand   \x   {\times}

\newcommand   \half{\frac 1 2}
\newcommand   \lptl{\raise .8ex\hbox{$^\leftarrow$} \hspace{-9pt} \partial} 
\newcommand   \lrptl{\raise .8ex\hbox{$^\leftrightarrow$} \hspace{-9pt} \partial} 
\newcommand \nn {\nonumber}
\def\be#1\ee{\begin{equation}\begin{aligned}#1\end{aligned}\end{equation}}
\makeatletter
\DeclareRobustCommand\bea{\@ifnextchar[{\@@bea}{\@bea}}
\def\@@bea[#1]#2\eea{\begin{subequations}\begin{align}#2\end{align}\label{#1}\end{subequations}}
\def\@bea#1\eea{\begin{subequations}\begin{align}#1\end{align}\end{subequations}}
\makeatother
\mathcode`l="8000
\begingroup
\makeatletter
\lccode`\~=`\l
\DeclareMathSymbol{\lsb@l}{\mathalpha}{letters}{`l}
\lowercase{\gdef~{\ifnum\the\mathgroup=\m@ne \ell \else \lsb@l \fi}}%
\endgroup
\renewcommand \a  {\alpha}
\renewcommand \b  {\beta}

\newcommand   \de {\delta}
\newcommand   \De {\Delta}
\newcommand   \e  {\epsilon}
\renewcommand \l  {\lambda}

\newcommand   \f  {\phi}

\newcommand   \bk {\mathbf{k}}

\newcommand   \bx {\mathbf{x}}

\newcommand   \R  {\mathbb{R}}

\newcommand   \so    {\mathrm{so}}

\newcommand   \cA {\mathcal{A}}

\newcommand   \cD {\mathcal{D}}

\newcommand   \cG {\mathcal{G}}

\newcommand   \cJ {\mathcal{J}}
\newcommand   \cK {\mathcal{K}}
\newcommand   \cL {\mathcal{L}}
\newcommand   \cM {\mathcal{M}}

\newcommand   \cO {\mathcal{O}}

\newcommand*{\diagArrowOne}{\rotatebox[origin=c]{135}{\(\Longleftrightarrow\)}}
\newcommand*{\diagArrowTwo}{\rotatebox[origin=c]{45}{\(\Longleftrightarrow\)}}
\begin{document}
\title{Momentum-space formulae for AdS correlators for diverse theories in diverse dimensions}

\def\andname{}
\author{Soner Albayrak$^{\text{\faEuro}\;,\;\text{\faGbp}\;,\;\text{\faTurkishLira}}$}
\author{Savan Kharel$^{\text{\faDollar}}$}
\author{Xinkang Wang$^{\text{\faDollar}}$}
\affiliation{$^{\text{\faEuro}}$Institute of Physics, University of Amsterdam, Amsterdam, 1098 XH, The Netherlands}
\affiliation{$^{\text{\faGbp}}$Center for Theoretical Physics, National Taiwan University, Taipei 10617, Taiwan}
\affiliation{$^{\text{\faTurkishLira}}$Department of Physics, Middle East Technical University, Ankara 06800, Turkey}
\affiliation{$^{\text{\faDollar}}$Department of Physics, University of Chicago, Chicago, IL 60637, USA}
\begin{abstract}
In this paper, we explore correlators of a series of theories in anti-de Sitter space: we present comprehensive results for interactions involving scalars, gluons, and gravitons in multiple dimensions. One aspect of our investigation is the establishment of an intriguing connection between the kinematic factors of these theories; indeed, such a connection directly relates these theories among themselves and with other theories of higher spin fields. Besides providing several explicit results throughout the paper, we also highlight the interconnections and relationships between these different theories, providing valuable insights into their similarities and distinctions. 
\end{abstract}
\date{\today}
\maketitle
\tableofcontents
\section{Introduction}
\label{sec:intro}

In the last decade, scattering amplitudes have been at the forefront of theoretical research. These developments have positioned scattering amplitudes at the vanguard of theoretical research, unlocking profound insights across various applications. These range from the obvious usage of making experimental predictions at the Large Hadron Collider (LHC) to innovative strides in the field of gravitational wave physics \cite{Elvang:2013cua, Buonanno:2022pgc}. In this period of prolific research, scientists have elucidated both basic tree-level and intricate multi-loop amplitude expressions \cite{Arkani-Hamed:2010zjl}. The hallmark of these formulations is their remarkable simplicity, contrasting sharply with the complexity typically encountered in intermediate computations of these structures. This observation has motivated researchers to produce explicit results, thereby amassing a substantial repository of ``theoretical data'' on scattering amplitudes. This endeavor aims to uncover numerous latent patterns and structures within scattering amplitudes. Indeed, this approach has proven exceptionally fruitful, leading to seminal developments, including the incorporation of positive geometry concepts and the pioneering discovery of the double copy principle \cite{Bern:2010ue, Arkani-Hamed:2017tmz}.

Another area of significant interest in contemporary theoretical physics is the holographic duality, particularly epitomized in the Anti-de Sitter/Conformal Field Theory (AdS/CFT) correspondence \cite{Maldacena:1997re,Witten:1998qj}. This duality intriguingly relates the boundary operators in a Conformal Field Theory (CFT) to bulk fields in Anti-de Sitter (AdS) space, offering a novel perspective to understand CFT correlation functions as scattering amplitudes in AdS space. Despite plethora of momentum basis results in the flat space scattering, momentum basis correlators in AdS remain less explored, partly due to the complexities of bulk integrals unique to AdS geometries. Nevertheless, the last decade has witnessed considerable progress in this domain, as evidenced by a range of studies \cite{Albayrak:2018tam, Albayrak:2019yve,  Albayrak:2019asr, Albayrak:2020isk, Albayrak:2020bso, Albayrak:2020fyp, Albayrak:2023jzl, Bzowski:2019kwd, Bzowski:2020kfw,Bzowski:2013sza,  Bzowski:2022rlz,Marotta:2022jrp,  Isono:2018rrb, Isono:2019wex, Coriano:2019nkw, Anand:2019lkt, Jain:2021wyn, Jain:2021vrv,  Armstrong:2023phb,  Farrow:2018yni, Coriano:2021nvn,Coriano:2020ees}.\footnote{For some position, Mellin, embedding space approaches, see for instance: \cite{Mack:2009mi,Penedones:2010ue,Rastelli:2016nze,Rastelli:2017udc,Alday:2020lbp,Alday:2020dtb,Alday:2022lkk, Alday:2021odx, Kharel:2013mka,Costa:2014kfa,Sleight:2017fpc,Goncalves:2019znr,Alday:2022lkk,Bissi:2022mrs,Li:2023azu,Alday:2023kfm,Chu:2023pea}}

This paper endeavors to make a contribution to the exploration of theories within the AdS/CFT framework, specifically in momentum space. It offers comprehensive explicit computations of correlators of several type of scalar and spinning fields, along with discussions of the interconnections among these diverse theories. Drawing inspiration from scattering amplitude studies, our exploration extends into various dimensions (see for instance  \cite{Cheung:2017ems}). We then conclude the paper with thoughtful speculation on potential avenues for future research.
\section{AdS Perturbation Theory in Momentum Space and Classification of Spinning Fields}
A Poincar\'e patch of a maximally symmetric homogeneous curved space can be described by the metric
\be 
ds^2=\frac{\rho^2}{z^2}\left(\pm dz^2+\eta_{ij}dx^idx^j\right)
\ee 
where $\rho$ (or $i\rho$) is the curvature radius and $\eta$ is the metric for the boundary at $z\rightarrow 0$. Depending on the signature of $\eta$ and the sign of $dz^2$, this metric describes Euclidean or Lorentzian
(A)dS. In the rest of the paper, we will mostly focus on AdS (hence $+dz^2$) and work with spacelike momenta, making all our computations agnostic to the signature of a mostly positive $\eta$. In addition, we will set $\rho=1$ for convenience.

The Poincar\'e coordinates are particularly advantageous due to the evident translational invariance in the boundary coordinates, denoted by $x^i$. This invariance simplifies analyses by allowing us to directly employ momentum space techniques for these coordinates, a method that aligns well with the established practices in the existing literature. Opting for this coordinate system offers significant benefits, which we underline as follows: \textbf{(a)} flat space limit is extremely simple \cite{Raju:2012zr}, making comparison with flat space data rather straightforward; \textbf{(b)} cosmological data is conventionally collected and stored in momentum coordinates, making comparison with dS literature far more intuitive; \textbf{(c)} diagrammatic rules are quite analogous with (and in some cases even exactly same to) the flat space diagrammatic rules, making this coordinate system an appealing choice for perturbation theory.

The steps in the derivation of perturbation theory rules in (A)dS are same with those of the flat space: we start with a weakly-coupled Lagrangian, derive the bulk to bulk propagator from the quadratic action, and derive the vertex factors from the interaction pieces. We will first illustrate these steps in case of scalars as a brief review, and then move on to the discussion of the spinning fields.

\subsection{Primer: scalar fields in curved background}
\label{sec:scalar review}
Consider the following action
\be 
\label{eq: scalar action in curved background}
\hspace*{-.5em}
S=\int d^{d+1}x\sqrt{g}\left[
\half\left(
g^{\mu\nu}(\partial_\mu\f)(\partial_\nu\f)+\mu^2\f^2
\right)+\cL_{\text{int}}(\f)
\right],
\ee 
where one is usually interested in polynomial interaction piece $\cL_{\text{int}}$ for its relevance to the cosmology though we will keep it generic.\footnote{
	The parameter $\mu^2$ is the \emph{effective mass square}, which is defined as $\mu^2\coloneqq m^2+\xi R$ for the Ricci scalar $R$ ---in AdS, $R=-d(d+1)/\rho^2$ \cite{Albayrak:2020isk}.} From the quadratic action, one can derive the equation for the \textit{bulk-to-bulk} propagator via the Green's method \cite{Raju:2011mp}:\footnote{We follow the same conventions as in \cite{Albayrak:2018tam, Albayrak:2019asr, Albayrak:2019yve, Albayrak:2020isk, Albayrak:2020bso, Albayrak:2020fyp, Albayrak:2023jzl}: $\bk$ denotes the boundary spacelike momenta with $\bk^2>0$, and $k$ its norm $\sqrt{\bk^2}$.}
\be 
\left(z^{d+1}\partial_z z^{1-d}\partial_z-z^2 \bk^2-\mu^2\right)\cG(\bk;z,z')=i\de(z-z')z^{d+1}.
\ee 
For the right hand side, we can use the identity
\be 
\int\limits_0^\infty pdp J_\b(pz)J_\b(p z')=\frac{\de(z-z')}{z},
\ee 
where $J_\beta$ is the Bessel function of the first kind. In the left hand side, we use the ansatz $\cG(\bk;z,z')=\int pdp\; c(p)(z z')^\a J_\b(pz)J_\b(pz')$. The differential operator then generates $(d-2\a)J_{\b-1}(pz)$ on the left-hand side, which we kill by setting $\a=d/2$. The rest of the equation fixes the unknown $\b$ and the function $c(p)$, leading to the following expression of bulk-to-bulk propagator
\be 
\label{eq: scalar bulk to bulk}
\cG(\bk;z,z')=\int\limits_0^\infty \frac{-ipdp}{p^2+\bk^2-i\epsilon}(z z')^{d/2} J_\nu(pz)J_\nu(pz')
\ee 
where we define the parameter $\nu$ as\footnote{This parameter is actually related to the scaling dimension $\De$ of the boundary operator that is dual to the scalar field: $\De=\nu+d/2$.}
\be 
\label{eq: representation in terms of effective mass}
\nu\coloneqq\frac{\sqrt{d^2+4\mu^2}}{2}. 
\ee 

One can now go ahead and derive \emph{bulk-to-boundary} propagator $\cG (\bk, z)$ by starting from the bulk-to-bulk propagator and taking one of the points to the boundary. However, the regularity in the bulk for spacelike momenta \cite{Raju:2011mp} and the compatibility with Ward identities already bootstraps the expression
\be 
\label{eq: scalar bulk to boundary}
\cG(\bk,z)=\sqrt{\frac{2}{\pi}}z^{d/2}k^{\nu}K_\nu(k z)
\ee 
where $K_\nu$ is the modified Bessel function of the second kind.\footnote{\label{footnote: normalization of spinning bulk to boundary}
	The overall factor $\sqrt{\frac{2}{\pi}}$ is chosen conventionally. When we move on to the spinning bulk fields dual to conserved non-scalar boundary operators, we will work with the generalization of this formula as 
	\be 
	\label{eq: bulk-to-boundary}
	\cG_{i_1\dots i_l}(\bk,z)=\epsilon_{i_1\dots i_l}\sqrt{\frac{2}{\pi}}z^{d/2-l}k^{\nu}K_\nu(k z)
	\ee 
	for the appropriate $\nu$.}

With the bulk-to-bulk and bulk-to-boundary propagators given as \eqref{eq: scalar bulk to bulk} and \eqref{eq: scalar bulk to boundary}, the only ingredient left to crank the perturbation theory machinery is the vertex factor: in the case of scalars with polynomial interactions (our focus in this paper), this is rather trivial ---the vertex factor is a simple constant. However, we will see non-trivial examples in the later sections where vertex factor will have a tensor part and a monomial $z-$dependence.

With all the ingredients derived, we can now consider the simplest computation: three-point Witten diagram of three generic scalar fields dual to scalar operators of arbitrary scaling dimensions $\De_a$. The amplitude is simply the product of bulk to boundary propagators integrated over the bulk point:
\begin{equation}
	\cA_3=  \int_0^\infty \frac{dz}{z^{d+1}} \prod_{a=1}^3 \left[z^{\frac{d}{2}} \sqrt{\frac{2}{\pi}}  k_a^{\nu_a} K_{\nu_a} (k_a z)\right]
\end{equation}
as seen for instance in \cite{Bzowski:2015pba}. By taking into account the appropriate normalization factors, one could also derive the three-point correlation function of the boundary CFT from this amplitude.\footnote{
	In principle, $\expval{\mathcal{O}_1 \mathcal{O}_2 \mathcal{O}_3 }=b_{\cO_1\cO_2\cO_3}\cA_3$ where $b_{\cO_1\cO_2\cO_3}$ contains the information of the conformal weights of the boundary operators, see eqn. 4.20 and 4.21 of \cite{Liu:2018jhs} as an example.
}

In extending our analysis to tree-level Witten diagrams with a four and higher number of points, the approach remains largely unchanged. the process still involves multiplying all the relevant propagator and vertex factors. Furthermore, an integration over all points in the bulk is necessary, incorporating the volume factor \( z^{-(d+1)} \). However, it is important to note that these higher-point functions might necessitate renormalization, as detailed in the work of Bzowski et al. \cite{Bzowski:2022rlz}. Moving forward, we encounter additional complexities when dealing with spinning fields, which will be the focus of the subsequent section.

\subsection{Classifications of spinning fields in AdS}
\label{sec:class}
In the preceding discussion, we outlined a methodology that can be  adapted for the analysis of spinning fields within the AdS framework. This adaptation involves a series of systematic steps:
\begin{enumerate}
	\item Derive the propagators from the quadratic action
	\item Derive vertex factors from the interaction pieces
	\item To compute any given tree-level Witten diagram, contract propagators and vertex factors and integrate them over the bulk points
\end{enumerate}

Although the almost-same prescription is well used for decades in the Minkowski space scattering amplitudes computations, there is actually a technical complication in AdS amplitudes that is absent in the flat space. We will discuss this novelty (and its implications) in this section, and turn to the physically relevant spinning fields in the next section. 

In flat space, the unitarity ensures that the internal legs of a  Feynamn diagrams takes the form 
\be 
\cG(x_1,x_2)=\int d^dp \;\frac{\sum\limits_j\e_j\e_j^\dagger}{p^2+m^2+i\epsilon}
e^{ip\.(x_1-x_2)}
\ee  
upto an overall phase, whereas the external legs are 
\be 
\cG(x)=\int d^dp \;\e_j e^{ip\.x}.
\ee 
Here $\e_j$ is a polarization vector with $j$ going from 1 to the number of physical degrees of freedom. Thus, any tree-level Feynman diagram can be brought to the form
\be 
\textrm{FD}=\int\prod_i d^dp_i\left[\frac{\prod_{i,j}c_{i,j}e^{ip_i\.x_j}}{\prod_i (p_i^2+m_i^2)}\right]\textrm{contraction}\left(V,\e^{(1)},\e^{(2)},\dots\right)
\ee 
for some scalars $c_{i,j}$, whereas $V$ denotes the interaction vertex.

A crucial property of the flat space is that the functional form of the \emph{kinematic factor} of the Feynman diagram, the term within square brackets above, does not explicitly depend on the spin of the fields or the dimension of the spacetime. However, this property does not extend to AdS: we can quickly check this by analyzing the form of the bulk-to-boundary propagator for a spin-$l$ gauge field in AdS$_{d+1}$ --see \equref{eq: bulk-to-boundary}.

This novelty in AdS brings a new challenge that is absent in the flat space computations: case by case calculation of kinematic factors for the Witten diagrams. Fortunately, one can actually classify these kinematic factors into universality classes: not only does this simplify the explicit computations, but can this also help with making connections between CFT correlators of various operators in different dimensions.

In what follows, let us restrict to theories with cubic interactions. We will take the interaction to be rather generic by inducing it via a piece in the Lagrangian that includes three arbitrary conserved fields of spin $l_1\ge l_2\ge l_3$ and $n$ derivatives.\footnote{In AdS$_4$, $l$ is the standard spin, i.e. the only representation label of $\so(3)$ group in the stabilizer subgroup of the AdS isometries. In higher dimensions, we refer the length of the first row of the Young tableaux of the $\so(d)$ representation as the spin.} The vertex factor due to such an interaction should have $l_1+l_2+l_3$ contravariant indices to contract with three legs (each with $l_1$, $l_2$, and $l_3$ indices) that enter into the vertex. In our position-momentum coordinate system in the Poincar\'e patch, $n$ of such indices are induced by the boundary momenta $\bk^i$ whereas remaining indices by $(l_1+l_2+l_3-n)/2$ many inverse boundary metrics $g^{ij}=z^{2}\eta^{ij}$. But since each momenta $\bk^i$ needs one inverse metric as well, we end up with the requirement of $(l_1+l_2+l_3+n)/2$ many inverse boundary metrics for such an interaction, which leads to the correspondence
\begin{figure*}
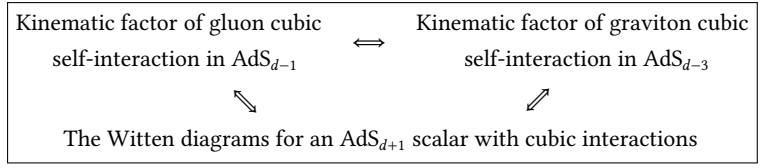

	$\boxed{
		\begin{aligned}
			&\begin{aligned}
				\text{Kinematic factor of gluon cubic}\\
				\text{self-interaction in AdS$_{d-1}$}\quad
			\end{aligned}\quad\Longleftrightarrow\quad
			\begin{aligned}
				&\text{Kinematic factor of graviton cubic}\\
				&\qquad\text{self-interaction in AdS$_{d-3}$}
			\end{aligned}
			\\
			&\qquad\qquad\qquad\qquad\quad\diagArrowOne\qquad\qquad\qquad\qquad\qquad\quad\diagArrowTwo
			\\
			&\qquad
			\text{The Witten diagrams for an AdS$_{d+1}$ scalar with cubic interactions}
		\end{aligned}
	}$
	\caption{\label{Figure} An example map of connections between the web of theories}
\end{figure*}
\begin{multline}
\hspace*{-1em}
\boxed{\text{cubic interaction with $n$ derivatives between }\cO_{l_1},\;\cO_{l_2},\;\cO_{l_3}}
\\\diagArrowOne\\
\boxed{z^{l_1+l_2+l_3+n}\text{ factor in bulk-point integration}}
\end{multline}
This argument extends to scalars as well; in fact, we can immediately write down
\begin{multline}
	\text{cubic interaction}\sim\int dz z^{\frac{d-2}{2}+n} E_{\nu_1}(k_1 z)E_{\nu_2}(k_2 z)E_{\nu_3}(k_3 z) \label{eq:KKJgen}\\
	\nu_i=\left\{\footnotesize\begin{aligned}
		l_i-2+d/2&\;\text{ conserved $l>0$ operators}\\\sqrt{d^2+4\mu^2}/2&\;\text{ scalars}
	\end{aligned}\right.
\end{multline}
where $E_\nu$ is Bessel $K_\nu$ or $J_\nu$. This is a generalization of several case studies in the literature.\footnote{For instance, graviton cubic self-interaction, gluon cubic self-interaction, scalar-scalar-graviton interaction, and scalar cubic nonderivative-interaction have $z^\#$ consistent with the general formula above, since these cases have respective $(n,l_1,l_2,l_3,\#)$ as $(2,2,2,2,8)$, $(1,1,1,1,4)$, $(2,2,0,0,4)$, and $(0,0,0,0,0)$.} We can summarize this observation as follows:
\begin{center}
	\itshape The kinematic factors of two theories in AdS$_{d}$ and AdS$_{d'}$ are the same if their perturbative spectrum and interaction Lagrangian are appropriately related!
\end{center}

The simplest implication of the result above is the correspondence in Figure~\ref{Figure}. Unsurprisingly, this shows that gluon and graviton kinematic factors can be re-interpreted as scalar Witten diagrams. But perhaps surprisingly, this also indicates that we can find an equivalence between gauge and graviton kinematic factors if we focus on appropriate spacetime dimensions.

By analyzing \eqref{eq: scalar action in curved background} we can obtain further correspondences; for instance,
\begin{subequations}
	\begin{multline}
	\boxed{	\text{$l_1-l_2-l_3$ interaction with $n-$derivative in AdS$_{d+1}$}}
		\\
		\diagArrowOne
		\\
		\boxed{
	\begin{aligned}
\text{$(l_1+1)-(l_2+1)-(l_3+1)$ interaction}\\
\text{with $(n+1)-$derivative in AdS$_{d-1}$}
	\end{aligned}	
	}
\end{multline}
	and
\begin{multline}
	\boxed{\text{$l_1-l_2-l_3$ interaction with $n-$derivative in AdS$_{d+1}$}}
		\\
		\diagArrowOne
		\\
\boxed{\begin{aligned}
\text{all scalars of $\De_i=d-2+l_i+n$ with}
\\
\text{polynomial interaction in AdS$_{d+2n+1}$}
\end{aligned}}
	\end{multline}
\end{subequations}

Although we present such relations, we will not dwell on their utility in detail in this paper. Indeed, higher spin theories are of particular importance in several contexts, most importantly in the Vasiliev theories; nevertheless, analysis of such theories would take us too far from our main point so we will stick to spin 0, 1, and 2 fields in the rest of the paper.\footnote{
	We note that AdS does allow consistent models with higher spin fields as the flat-space no-go theorems are no longer valid here \cite{Giombi:2016ejx}.
}

\subsection{Pure Yang-Mills and Einstein gravity}
Consider Yang-Mills action in curved background
\begin{subequations}
	\label{eq: pure YM and gravity}
	\be 
	\label{eq: spin1 action in curved background}
	S=- \frac{1}{4} \int d^{d+1} x \sqrt{g} F_{\mu \nu}^a F^{\mu \nu, a}
	\ee 
	where $F_{\mu \nu}^a = \nabla_\mu A_\nu^a - \nabla_\nu A_\mu^a + e f^{abc} A_{\mu}^b A_{\nu}^c$ with the coupling constant $e$. Likewise, the usual Einstein-Hilbert bulk action with a cosmological constant term reads as
	\be 
	\label{eq: spin2 action in curved background}
	S = \int d^{d+1} x \sqrt{g} (R[g] -  2 \Lambda). 
	\ee 
\end{subequations}
Since we are interested in doing a well-defined perturbation expansion in AdS background, we will take $e$ small, $\Lambda = - \frac{d(d+1)}{2}$, and $g=g_{\text{AdS}}+h$ for small $h$: this leads to the following well-known bulk-to-boundary propagators
\bea 
\cG_i^a(\bk,z)=&\sqrt{\frac{2}{\pi}}\e_i^a(\bk)z^{d/2-1}k^{d/2-1}K_{d/2-1}(kz) \label{eq:gluonbtoB}
\\
\cG_{ij}(\bk,z)=&\sqrt{\frac{2}{\pi}}\e_{ij}(\bk)z^{d/2-2}k^{d/2}K_{d/2}(kz) \label{eq:gravitonbtoB}
\eea  
which are compatible with the general result in \eqref{eq: bulk-to-boundary}. We only present the boundary components of $\cG_\mu$ and $\cG_{\mu\nu}$ as we are working in the \emph{axial gauge}, see \cite{Raju:2011mp} for further details. This also means that the \emph{polarization tensors} $\e_i$ and $\e_{ij}$ are practically identical to their flat space counterparts in $\R^d$: they are transverse, null, and traceless. Whenever there is no room for confusion, we will drop their explicit momentum dependence ---same with the color indices of $\cG_i^a$.

The derivation of bulk-to-bulk propagators of gluon and graviton is analogous to our derivation of scalar propagator in \S~\ref{sec:scalar review}: after some computation, we end up with
\bea 
\mathcal{G}_{ij}(\bk;z_1,z_2) =&  \int_0^\infty \frac{-ipdp(z_1 z_2)^{\nu}}{p^2 + \textbf{k}^2  - i \epsilon} J_{\nu} (p z_1) J_{\nu} (p z_2) H_{ij}(\bk,p)
\label{eq:gluonbtob}\\
\mathcal{G}_{ij,kl}(\bk;z_1,z_2) =&  \int_0^\infty \frac{-ipdp(z_1 z_2)^{\nu-2}}{ p^2 + \textbf{k}^2 - i \epsilon} J_{\nu} (p z_1) J_{\nu} (p z_2) H_{ij,kl}(\bk,p) \label{eq:gravitonbtob}
\eea  
where $\nu=d/2-1$ for gluon and $\nu=d/2$ for graviton. Here, the tensor structures $H$ are defined as 
\bea[eq:propagator tensor structures]
H_{ij}(\bk,p) \coloneqq &\eta_{ij} + \frac{\textbf{k}_i \textbf{k}_j} {p^2}\label{eq: definition of H}
\\
H_{ij,kl}(\bk,p) \coloneqq&H_{ik}(\bk,p) H_{jl}(\bk,p) + H_{il}(\bk,p) H_{jk}(\bk,p)\nn\\{}&- \frac{2}{(d-1)} H_{ij}(\bk,p) H_{kl}(\bk,p).
\eea

From the non-quadratic part of the actions in \eqref{eq: pure YM and gravity}, one can derive the vertex factors for the color ordered YM amplitude and the cubic gravitational interaction:
\bea[eq:vertex factors for YM and Gravity]
V^{ijk}_{\bk_1,\bk_2,\bk_3}\coloneqq{}&
\frac{i z^4}{\sqrt{2}}
\left(
\eta^{ij}(\bk_1-\bk_2)^k+\eta^{jk}(\bk_2-\bk_3)^i+\eta^{ki}(\bk_3-\bk_1)^j\right)
\\
V^{ijkl}\coloneqq{}&{} \frac{i z^4}{2} \left(2\eta^{ik} \eta^{jl}-\eta^{ij} \eta^{kl}-\eta^{il} \eta^{jk}\right)\;,
\\
V^{ijklmn}_{\bk_1,\bk_2,\bk_3}\coloneqq{}&{} \frac{z^8}{4}\left[
\left(
\bk_2^i\bk_3^j\eta^{km}\eta^{ln}-2\bk_2^i\bk_3^k\eta^{jm}\eta^{ln}
\right)+\text{ permutations}
\right]
\eea
See \cite{Albayrak:2020bso} for further details.

\subsection{Orthogonal decomposition of gauge and gravity propagators}
\label{sec:orthdecom}
The tensor structures of gauge and gravity propagators are constructed from the object $H$, as defined in \eqref{eq: definition of H}. This object can be elegantly decomposed into the projectors $\widehat{P}_{ij}^{\perp}$ and $\widehat{P}_{ij}^{\parallel}$ , adhering to standard projection rules: $P^\perp+P^\parallel=\de$ and $P^\a\.P^\b=\de^{\a\b}P^\a$. Explicitly,
\be\label{eq:projectors}
\widehat{P}_{ij}^{\perp}(\bk)  \coloneqq {} \eta_{ij} - \frac{\textbf{k}_i \textbf{k}_j}{k^2}\;,\quad
\widehat{P}_{ij}^{\parallel}(\bk)  \coloneqq {} \frac{\textbf{k}_i \textbf{k}_j}{k^2}.
\ee 
Clearly, the tensor $H_{ij}$ of gluon decomposes as
\begin{subequations}
	\be 
	H_{ij}(\bk, p)  = \widehat{P}_{ij}^{ \perp}(\bk) + \frac{(k^2 +p^2)}{p^2}\widehat{P}_{ij}^{\parallel}(\bk) 
	\ee 
	whereas the decomposition of $H_{ij,kl}(\bk, p)$ of graviton is best written as\footnote{See \cite{Albayrak:2019yve} for further details on this decomposition. A similar decomposition with a different set of conventions can also be seen in \cite{Baumann:2019oyu}.
	}
	\begin {multline}
	H_{ij,kl}(\bk, p)=  \frac{(k^2 + p^2)} {p^2}  \left(\widehat{P}_{ij, lm}^{(\perp, \parallel)}(\bk) + \widehat{P}_{ij, lm}^{(\parallel, \perp)}(\bk) \right)\\+\widehat{P}_{ij, lm}^{(\perp, \perp)}(\bk)+ \left(\frac{k^2 + p^2}{p^2}\right)^2 \widehat{P}_{ij, lm}^{(\parallel, \parallel)}(\bk),
	\end {multline}
\end{subequations}
where we have defined 
\begin {align}
\widehat{P}_{ij, lm}^{(\alpha, \beta)} \coloneqq  \widehat{P}_{il}^{(\alpha)} \widehat{P}_{jm}^{(\beta)} + \widehat{P}_{im}^{(\alpha)} \widehat{P}_{jl}^{(\beta)} - \frac{2}{d-1} \widehat{P}_{ij}^{(\alpha)} \widehat{P}_{lm}^{(\beta)}.
\label{eq:Gpro}
\end {align}
One can now decompose the gauge and gravity propagators into their \emph{transverse} and \emph{longitudinal} parts. For brevity, we choose the convention of defining $\widehat{P}_{ij, lm}^{(\perp, \perp)}$ part of the graviton propagator as its transverse part, and the rest as its longitudinal part.

\subsection{Interactions among scalars, gluons, and gravitons}
\label{sec:int}
In preceding discussions, we methodically derived the propagators pertinent for gluons, gravitons, and scalars with an arbitrary effective mass; we also presented the vertex factors for gluon and graviton-only interactions in \eqref{eq:vertex factors for YM and Gravity} (recall that the vertex factor for scalar-only interaction is simply a constant). This section aims to delve into the intricate dynamics of inter-particle interactions, thereby laying a foundational framework for the ensuing discourse in this paper.

Consider the simplest self-consistent theory with non-trivial interaction pieces, i.e. scalar-QCD (and its special case scalar-QED) with the action
\begin{multline}
S= \sum\limits_{i=1}^{N_f}\int d^{d+1}x\sqrt{g}\bigg(
g^{\mu\nu}\left(\cD_\mu \Phi_{i} \right)^\dagger \cD_\nu\Phi_{i}
\\
+\mu^2_i \Phi_{i}^\dagger \Phi_{i}+ \frac{1}{2}\mathrm{tr}\left(F_{\mu \nu} F^{\mu \nu}\right) \bigg) 
\label{eq:sQCDaction}
\end{multline}
for the gauge covariant derivative $\cD$ and the field stress tensor $F_{\mu\nu}\coloneqq[\cD_\mu,\cD_\nu]/(ie)$. The action is written for $N_f$ flavors of scalars with a diagonalized mass matrix ---hence $(\mu^2)_{ij}\Phi_i^\dagger\Phi_j=\mu^2_i \Phi_{i}^\dagger \Phi_{i}$; we also suppressed the color indices in $\Phi_i$, which can be in any representation of the color group. Finally, we note that $A^\dagger B$ refers to the singlet in the branching of the product representation $A^\dagger\otimes B$; usually, one considers the fundamental representation (hence a vector $\Phi$) or the adjoint representation (hence a matrix $\Phi$).

Let us proceed with a scalar field $\Phi_i$ in the fundamental representation of the color group; we can then immediately write down
\be 
(\cD_\mu\Phi_i)_a=\nabla_\mu\Phi_{i,a}+ie(A_\mu)_a^{\;\;b}\Phi_{i,b}
\ee 
for the color indices $a$ \& $b$, where $A$ is in the adjoint representation as usual and $\nabla$ is the standard spacetime covariant derivative. By inserting this in the action above and expanding in the small parameter $e$, we can extract several leading-order interaction Lagrangians from which the relevant vertex factors for the perturbation theory can be derived; for instance, \emph{gluon emission of a scalar} can be derived from
\be 
S_{\text{int}} = i e \sum\limits_{i=1}^{N_f}\int d^{d+1} x \sqrt{g}\left(\Phi_{i,a}^*\overleftrightarrow{ \nabla_\mu} \Phi_{i,b}\right) (A^{\mu})^{ab}. 
\label{eq:gluonint}
\ee 
where $A\overleftrightarrow{\partial}B\equiv A\partial B-(\partial A)B$ as usual. Furthermore one can also derive higher point interactions by inserting $g=g_{\text{AdS}}+h$ and expanding for small $h$.

The next theory we will analyze is that of a scalar in a \emph{dynamical} background; as a simple model, consider the action
\begin{multline}
S=\int d^{d+1}x\sqrt{g}\bigg(
-\half\left(
g^{\mu\nu}(\partial_\mu\f)(\partial_\nu\f)+m^2\f^2
\right)
\\
+ \frac{2-\xi\f^2}{2}R[g]-2 \Lambda
\bigg)
\end{multline}
and expand it around AdS, i.e. $g=g_{\text{AdS}}+h$ for small $h$; as before, we can get interesting interaction Lagrangians. As an example, \emph{graviton emission of a scalar} can be derived from
\be 
\label{eq: graviton scalar interaction}
S_{\text{int}} = \int d^{d+1} x \sqrt{g}  h^{\mu \nu}  \left((\partial_\mu \phi)(\partial_\nu \phi) - \frac{1}{2} g_{\mu \nu} g^{\alpha \beta}(\partial_{\alpha} \phi) (\partial_{\beta} \phi)  \right)
\ee 
where the metric $g$  is taken to be the background metric in the leading order.

In the following sections, we will discuss such interactions in greater depth.\footnote{We would like to note that the name of this subsection is not meant to imply a complete discussion of all possible interactions: we only introduced the cases which we will utilize in the rest of the paper. For instance, one could add a cubic interaction between two different flavors of scalars with different $U(1)$ charges in a scalar-QED, i.e.
	\be 
	S_{\text{int}}=\lambda\int d^{d+1}x\sqrt{g}\;\Phi\Phi(\Phi')^*
	\ee 
	for $q(\Phi')=2q(\Phi)$ where $q$ denotes the $U(1)$ charge.
	
}

\section{Explicit Computations in AdS$_{d+1}$}
In this section, our investigation expands into various AdS theories of physical significance, and compute contributions to the four-point correlation functions at leading order in the perturbation theory. 
Our analysis is distinguished by its breadth, covering purely scalar theories, scalar-QED, scalar gravity, and pure Yang-Mills gauge theory, and it spans across diverse dimensions. We leverage the orthogonal decomposition formalism, delineated in Section \ref{sec:orthdecom}, as our computational framework.

\subsection {Scalar-scalar interaction}
\label{sec:scsc}
Let us start by considering the following two-derivative scalar interaction action
\begin {align}
S_{\text{int}}^{\Phi, \Psi} = \int_{\text{AdS}} d^{d+1} x \sqrt{g} (\nabla \Phi)^2 \Psi
\label{eq:ssaction}
\end {align}
where $\Psi$ is viewed as a perturbative scalar field (e.g., $\Psi$ absorbs the weak coupling constant). The contribution to the correlation function of external $\Phi$'s due to the exchange of an internal $\Psi$, denoted as $ \big\langle \Phi (\textbf{x}_1) \Phi (\textbf{x}_2) \Phi ( \textbf{x}_3) \Phi (\textbf{x}_4) \big\rangle_\Psi$, is equivalent to the four-point Feynman diagram in leading order:
\begin {multline}
\big\langle \Phi (\textbf{x}_1) \Phi (\textbf{x}_2) \Phi ( \textbf{x}_3) \Phi (\textbf{x}_4) \big\rangle_\Psi=
\int \frac{d z dz' d\textbf{x}^d d\textbf{x}'^d} {z^{d+1} z'^{d+1}}
\\\x\left(g^{\mu \nu} \partial_\mu \cG^{\Phi}(z,\bx;\bx_1) \partial_\nu \cG^{\Phi}(z,\bx;\bx_2) \right)\mathcal{G}^\Psi (z, \textbf{x}; z', \textbf{x}')\\\x\left(g^{\delta \lambda} \partial_\delta \cG^{\Phi}(z',\bx';\bx_3) \partial_{\lambda} \cG^{\Phi}(z',\bx';\bx_4) \right) 
\label{eq:scalaronshell}
\end {multline}
where $g_{\mu \nu}$ is the AdS Poincar\'e metric and the volume factor $z^{-(d+1)}$ comes from the metric determinant $g$. Note that $\cG^\Phi$ are the bulk-to-boundary propagators of scalar $\Phi$ with points $\textbf{x}_a$ on AdS$_{d+1}$ boundary, while $\cG^\Psi$ is the bulk-to-bulk propagator of scalar $\Psi$. 

Although it is rather straightforward to generate the leading order correlator in the position space, we are usually interested in momentum-space expressions, not least because of their relevance for the observational data. The translational invariance at the boundary dictates an overall momentum conservation. Furthermore, note that \eqref{eq:scalaronshell} contains two distinct structures 
\bea 
g^{zz} (\partial_z \cG^{\Phi}) (\partial_z \cG^{\Phi}) \rightarrow{}&{} z^2 r_{\nu_\Phi} (k_a, k_b, z) \cG^{\Phi}(z, \bk_a) \cG^{\Phi}(z, \bk_b) \label{eq:gzz}\\
g^{ij} (\partial_i \cG^{\Phi}) (\partial_j \cG^{\Phi}) \ \rightarrow{}&{}  - z^2 \textbf{k}_a \cdot \textbf{k}_b \cG^{\Phi}(z, \bk_a) \cG^{\Phi}(z, \bk_b), \label{eq:gii}
\eea 
where the function $r_{\nu}$ on the first line is
\be 
\bigg( \frac{d-2\nu}{2 z} - k_a \frac{K_{\nu-1} (k_az)} {K_{\nu} (k_az)}\bigg) \bigg( \frac{d-2\nu}{2 z} - k_b \frac{K_{\nu-1} (k_bz)} {K_{\nu} (k_bz)}\bigg).
\ee 
Therefore, we have a mixture of terms with different $z$-powers, and the highest $z$-power term comes from \eqref{eq:gii} as well as from the ratio of Bessel $K$ functions within $r_\nu$. In the final expression of the correlator, the highest $z$-power term contributes to the leading order total-energy structure, which is relevant in the flat-space limit, while the lower $z$-power terms are correction terms to the subleading total-energy structures. We demonstrate below the evaluation of the highest $z$-power contribution in the correlator. 

Let us first consider the contribution of \eqref{eq:gii} in \eqref{eq:scalaronshell}. We can conveniently extract a pair of momentum inner product $(\bk_a\.\bk_b)$ appropriate for the channel we are interested in: this leads to the following $s$-channel expression
\begin{multline}
\label{eq:scalargii}
(2\pi)^d \delta^d \bigg(\sum_{a=1}^4 \textbf{k}_a \bigg) \x (\textbf{k}_1 \cdot \textbf{k}_2) (\textbf{k}_3 \cdot \textbf{k}_4) \mathcal{M}_{\perp; \ 2}^{\nu_\Phi,\nu_\Phi,\nu_\Phi,\nu_\Phi;\nu_\Psi} 
\end{multline}
for the ``seed'' amplitudes
\bea[eq:Mboth]
\hspace*{-.7em}
\mathcal{M}_{\perp; \ n}^{\nu_1,\nu_2,\nu_3,\nu_4;i} \coloneqq & \int_0^\infty \frac{pdp}{s^2 + p^2}\cK\cK\cJ^{(n)}_{\nu_1,\nu_2,i} (k_1, k_2, p)\cK\cK\cJ^{(n)}_{\nu_3,\nu_4,i} (k_3, k_4, p) \label{eq:Mtran}
\\\hspace*{-.7em}
\mathcal{M}_{\parallel; \ n}^{\nu_1,\nu_2,\nu_3,\nu_4;i}\coloneqq& \int_0^\infty \frac{dp}{p}\cK\cK\cJ^{(n)}_{\nu_1,\nu_2,i} (k_1, k_2, p)\cK\cK\cJ^{(n)}_{\nu_3,\nu_4,i} (k_3, k_4, p), \label{eq:Mlong}
\eea
where $s = \abs{\bk_1+\bk_2}$ and the function $\cK\cK\cJ$ is defined in accordance with \eqref{eq:KKJgen} and \eqref{eq: bulk-to-boundary}:
\be 
\label{eq: definition of KKJ}
\cK\cK\cJ^{(n)}_{\nu_a, \nu_b, i}\coloneqq \frac{2 k_a^{\nu_a}k_b^{\nu_b}}{\pi} \int_0^\infty z^{\frac{d-2}{2}+n}dz K_{\nu_a} (k_a z) K_{\nu_b} (k_b z) J_{i} (p z) \ .
\ee
Note that the number of derivatives is $n=2$ in the action \eqref{eq:ssaction}. Although the second line \eqref{eq:Mlong} does not appear in the scalar four-point function \eqref{eq:scalargii}, it is shown here for completeness, and as we will show later that it appears in the longitudinal components of four-point functions in scalar-QED and other theories (as the subscript already indicates, the first line \eqref{eq:Mtran} would appear in the corresponding transverse components).

On the other hand, the contribution of \eqref{eq:gzz} in \eqref{eq:scalaronshell} depends on the specific value of $\nu_\Phi$. Let us thus examine a simple scalar interaction theory in which the two scalars $\sigma$ and $\f$, with their own quadratic actions given as \eqref{eq: scalar action in curved background}, have effective mass squares $\mu_\sigma^2=1-d$ and $\mu_\f^2=0$, respectively. The interaction of the form $\int (\nabla \sigma)^2 \phi$ then yields the following four-point correlator in $s$-channel\footnote{Other channels can be obtained by a simple permutation of the momenta.}
\begin{multline}
\label{eq:scalarexch}
\big\langle \sigma (\textbf{k}_1) \sigma (\textbf{k}_2) \sigma (\textbf{k}_3) \sigma (\textbf{k}_4) \big\rangle_{\phi}^{(s)} = 
(2\pi)^d \delta^d \bigg(\sum_{a=1}^4 \textbf{k}_a \bigg) \\\x (k_1 k_2 - \textbf{k}_1 \cdot \textbf{k}_2) (k_3 k_4 - \textbf{k}_3 \cdot \textbf{k}_4) \mathcal{M}_{\perp; \ 2}^{\nu_\sigma,\nu_\sigma,\nu_\sigma,\nu_\sigma;\nu_\phi} + c.t.
\end{multline}
where the correction terms due to lower $z$-powers are denoted as ``$c.t.$".  Comparing with \eqref{eq:scalargii}, it is evident that the highest $z$-power contribution from \eqref{eq:gzz} generates momentum scalar products $(k_ak_b)$. 

Let us look at two specific dimensions as examples. At $d=3$, the interaction of the form $\int (\nabla \sigma)^2 \phi$ would then lead to
\be 
\cK\cK\cJ_{\nu_\sigma, \nu_\sigma, \nu_\phi}^{(2)} (k_1, k_2, p) =\sqrt{\frac{2}{\pi}} \frac{2 p^{\frac{3}{2}}}{(k_{12}^2 + p^2)^2}, 
\ee 
where we have used a shorthand notation $k_{ab} = k_a + k_b$. At this dimension, $\sigma$ and $\f$ are commonly referred to as conformally coupled scalar and massless scalar in literature, respectively, with $\Delta_\sigma =2$ and $\Delta_\phi=3$. Using the above $\cK\cK\cJ$ expression, we then obtain the following evaluation for correlator $\langle \sigma \sigma \sigma \sigma \rangle_{\phi}^{(s)}$ in AdS$_4$
\begin {multline}
(2\pi)^3 \delta^3 \bigg(\sum_{a=1}^4 \textbf{k}_a \bigg) \frac{2 (E_L E_R + s E)} {E_L^2 E_R^2 E^3} \\
\times  (k_1 k_2 - \textbf{k}_1 \cdot \textbf{k}_2) (k_3 k_4 - \textbf{k}_3 \cdot \textbf{k}_4) + c.t.
\end {multline}
where $E_{L,R}$ are left/right partial energy sums and $E$ is the total energy.\footnote{
In terms of momenta, these variables read as 
\be 
E=k_1+k_2+k_3+k_4\;,\quad E_L=s+k_1+k_2\;,\quad E_R=s+k_3+k_4.
\ee 
} We can straightforwardly check that we have the correct flat-space limit as well, which amounts to obtaining the coefficient of the leading total-energy pole when $E$ approaches zero \cite{Maldacena:2011nz,Raju:2012zr}, with which we arrive at $S$ for four-dimensional Mandelstam variable $S\coloneqq (k_1+k_2)^2 - s^2$.\footnote{The other two Mandelstam variables are
\begin {align}
T \coloneqq (k_1+k_4)^2 - t^2 , \quad U \coloneqq (k_1 + k_3)^2 - u^2
\end {align}
where $ t = |\textbf{k}_1 + \textbf{k}_4|$ and $u =|\textbf{k}_1 + \textbf{k}_3|$ (and $s = |\bk_1 + \bk_2|$) are the exchange momenta in respective channels. Therefore, the total four-point amplitude $\mathcal{A} = S + T + U$ is a constant. } For the same pair of interactive scalars, we can repeat the process to obtain the correlator expression in AdS$_6$ \footnote{The $R_n$ functions in the seed amplitude are defined as
\bea 
R_5 =& 12 k_1 k_2 k_3 (k_1 + k_2 + k_3) (k_{12} - s)^2 E_L^2 E_R^3 \\
R_4 =& 3 E_L E_R^3 (k_{12} - s) \big[ k_1 k_2 k_{12} (k_{12}^2 - s^2)\nn\\& + k_3 \big(k_{12}^4 - (k_1 - k_2)^2 s^2 \big) + k_{12} k_3^2 (k_{12}^2 + 4k_1k_2 - s^2) \big] \\
R_3 =& \big[k_{12}^6 - s^2 k_{12}^4 + s^2 (k_1 - k_2)^2 (s^2 - k_{12}^2) \nn\\&+\big(2s^2 k_{12} k_3 + k_3^2 (3k_{12}^2 - s^2)\big) (k_{12}^2 + 4k_1 k_2 - s^2)\big] E_R^3 \\
R_2 =& (k_{12}^2 + 4k_1 k_2 - s^2) \big[s^2 (k_1 + k_2 + k_3) + 3 k_{12} k_3^2 \big] E_R^3 \\
R_1 =& (k_{12}^2 + 4k_1 k_2 - s^2) (s^2 + 3k_3^2) E_R^3 \\
R_0 =& (k_{12}^2 + 4k_1 k_2 - s^2) \big[- 2 s^2 k_3 (s+ k_3) - s (s^2 + sk_3 + 3k_3^2)E_R \nn\\&- (s^2 + 3k_3^2)E_R^2 \big].
\eea 
}
\begin{multline}
(2\pi)^5 \delta^5 \bigg(\sum_{a=1}^4 \textbf{k}_a \bigg) \frac{2}{E_L^3 E_R^3 (k_{12} - s)^3} \bigg(\sum_{n=0}^5 \frac{R_n}{E^n} \bigg) \\
\times  (k_1 k_2 - \textbf{k}_1 \cdot \textbf{k}_2) (k_3 k_4 - \textbf{k}_3 \cdot \textbf{k}_4) + c.t.
\end {multline}
which then expectedly leads to the same flat-space amplitude as the AdS$_4$ correlator (up to a pre-factor of $k_1 k_2 k_3 k_4$).

\subsection{Scalar-QED}
The procedure of obtaining the contributions to the correlation functions at the leading order in perturbation theory remains the same when we move from scalar-only interaction to the scalar-QED. We start with the relevant interaction action, which in this case is
\be 
S_{\text{int}}^{\text{sQED}} = e \int_{\text{AdS}} d^{d+1} x \sqrt{g} \big({\tl\Psi} \nabla_\mu \Psi - \Psi \nabla_\mu {\tl\Psi}) A^\mu,
\ee 
where $\Psi$ and $\tl\Psi$ are real and imaginary parts of the complex scalar field respectively.\footnote{
Explicitly, we define
\be 
\Psi = \frac{\big(\Phi + \Phi^*\big)}{\sqrt{2}} , \quad \tl\Psi = \frac{\big(\Phi - \Phi^*\big)}{i \sqrt{2}}
\ee
where $\Phi$ and $\Phi^*$ are complex scalar fields. }
We will now proceed with the actual computation of the Feynman diagrams; however, we would like to specialize in two physically relevant cases in what follows, namely
\bea 
\text{conformally coupled scalar: }&\nu_\varphi =\frac{1}{2}\qquad\rightarrow\;\De_\varphi =\frac{d+1}{2} \\
\text{vectorlike scalar: }&\nu_\sigma =\frac{d-2}{2}\;\rightarrow\;\De_\sigma =d-1
\eea
where $\nu$ is the index that appears in the propagators, and $\De$ is the scaling dimension of the boundary operator that is dual to the relevant bulk field. Although these two cases collapses into one at $d=3$, the most commonly analyzed dimension in the literature, they physically describe different scalars in general dimensions: $\nu=1/2$ (along with the additional masslessness condition) makes the theory \eqref{eq: scalar action in curved background} perturbatively invariant under conformal transformations, whereas $\nu=(d-2)/2$ makes the scalar mimic the scaling behavior of the spin$-1$ bulk fields as we already discussed within scalar-scalar interactions.

Conformally coupled scalars have received extensive attention in the literature, owing to their relevance and inherent simplicity. Indeed, their straightforward nature not only facilitates deeper understanding but also allows them to serve as a foundational basis for computing more complex theories, as discussed in \cite{Baumann:2019oyu, Albayrak:2020isk}; in fact, their computations are almost identical to ordinary field theoretical computations in half flat spacetime.\footnote{This follows from the nice behavior of the conformally coupled scalar under a Weyl transformation that relates the metrics $ds^2=\frac{dz^2+dx_i dx^i}{z^2}$ and $ds^2=dz^2+dx_i dx^i$.} On the contrary, vectorlike scalars with $\De=d-1$ which are not conformally invariant are not analyzed as much;\footnote{The exception is $d=3$ at which vectorlike scalars coincide with the conformally coupled scalars and hence have been extensively studied.} nevertheless, due to the common scaling dimension, it becomes rather interesting to study the interaction of vectorlike scalars and spin-1 gauge fields, an interesting special case of scalar QED.

For both conformally coupled and vectorlike scalars, we can immediately write down the perturbative expression for the correlation function in the leading order; explicitly,
\begin {multline}
\hspace*{-1.5em}
\<\Psi(\bx_1)\tilde{\Psi}(\bx_2)\Psi(\bx_3)\tilde{\Psi}(\bx_4)\>_A = \int \frac{d z dz' d\textbf{x}^d d\textbf{x}'^d} {z^{d+1} z'^{d+1}}\mathcal{G}_{il} (z, \textbf{x}; z', \textbf{x}')\\\x
\left(g^{ij} \cG^{\tl\Psi}(z,\bx;\bx_1) \partial_{j} \cG^{\Psi}(z,\bx;\bx_2)-(\Psi\leftrightarrow\tl\Psi) \right) \\\x\left(g^{lm}\cG^{\tl\Psi}(z',\bx';\bx_3) \partial_{m} \cG^{\Psi}(z',\bx';\bx_4) -(\Psi\leftrightarrow\tl\Psi)\right)
\label{eq:scalargluon}
\end {multline}
which is the contribution to the four-point correlator due to the photon exchange.\footnote{We remind the reader that we choose the axial gauge throughout the paper, hence $A^z=A^0=0$ and there is no $g^{zz}$ contribution. } We will be explicitly working in the $s$-channel, but the $t$- and $u$- channels can be obtained via an appropriate permutation of the indices. We can now go ahead and decompose this correlator into transverse and longitudinal parts. In momentum space, the correlator expression for conformally coupled scalars reads as 
\begin{multline}
\big\langle \varphi (\textbf{k}_1) \tilde{\varphi} (\textbf{k}_2) \varphi (\textbf{k}_3) \tilde{\varphi}(\textbf{k}_4) \big\rangle_{A, \a}^{(s)} = (2\pi)^d \delta^d \bigg(\sum_{a=1}^4 \textbf{k}_a \bigg) \\\x(\textbf{k}_1 - \textbf{k}_2)^i (\textbf{k}_3 - \textbf{k}_4)^j \widehat{P}_{ij}^{\a} (\textbf{k}_1+\textbf{k}_2) \mathcal{M}_{\a; \ 1}^{\nu_\varphi, \nu_\varphi, \nu_\varphi, \nu_\varphi; \nu_A}
\end{multline}
where $\a=\perp,\parallel$ and the projectors $\widehat{P}_{ij}^{\a}$ are defined in \equref{eq:projectors}, while the seed amplitudes $\cM_\alpha$ are defined in \eqref{eq:Mboth}. Note that one can obtain the same set of expressions for vectorlike scalars by simply switching the index $\nu_\varphi$ to $\nu_\sigma$ in the above expressions.

We now present the explicit evaluation for $d=3$ and $d=5$ below. Even though one can also write down implicit results in general dimensions, explicit results are harder to obtain as the expression in generic dimensions is an integral of products of hypergeometric functions. After a little bit algebra, we can show that the full correlator in AdS$_{4}$ reads
\be 
\langle \varphi \tilde{\varphi} \varphi \tilde{\varphi} \rangle_{A, \perp}^{(s)}\evaluated_{d=3} & = \frac{(\textbf{k}_1 - \textbf{k}_2) \cdot (\textbf{k}_3 - \textbf{k}_4) + s^{-2} (k_1^2 - k_2^2) (k_3^2 - k_4^2)}{E_L E_R E} \\
\langle \varphi \tilde{\varphi} \varphi \tilde{\varphi} \rangle_{A, \parallel}^{(s)}\evaluated_{d=3} & = - \frac{(k_1 - k_2) (k_3 - k_4) }{s^2 E}, \label{eq:scalarqedd3}
\ee 
upto the prefactor $(2\pi)^3 \delta^3 \big(\sum_{a=1}^4 \textbf{k}_a \big)$.\footnote{
We remind the reader that these results are applicable both for conformally coupled and vectorlike scalars as they coincide in AdS$_4$.
} It is straightforward to check that the full correlator expression above is consistent with the corresponding flat-space amplitude. Indeed, in the flat-space limit $E \rightarrow 0$, the coefficient of the total-energy pole in the correlator $\langle \varphi \tilde{\varphi} \varphi \tilde{\varphi} \rangle_{A}^{(s)}$ becomes the corresponding flat-space scattering amplitude 
\begin {align}
\mathcal{A}_{\varphi \tilde{\varphi} \varphi \tilde{\varphi}}^{(s)} \evaluated_{d=3} & = \lim_{E \rightarrow 0} \frac{(\textbf{k}_1 - \textbf{k}_2) \cdot (\textbf{k}_3 - \textbf{k}_4) - (k_1 - k_2) (k_3 - k_4)} {E_L E_R},
\label{eq:flatD4}
\end {align}
which is simply $2S^{-1} (U- T)$, where we have imposed conservation of momentum and $S, T, U$ are the four-dimensional flat-space Mandelstam variables defined earlier. The same set of momentum-space expressions can be derived similarly in AdS$_6$. For the conformally coupled scalar, we have
\bea 
\langle \varphi \tilde{\varphi} \varphi \tilde{\varphi} \rangle_{A, \perp}^{(s)} \evaluated_{d=5} = & \frac{ (\textbf{k}_1 - \textbf{k}_2) \cdot (\textbf{k}_3 - \textbf{k}_4) + s^{-2} (k_1^2 - k_2^2) (k_3^2 - k_4^2) }{E_L^2 E_R^2} 
\nn\\
&\x \frac{2\big(E_L E_R + s E\big)}{E^3}
\\
\langle \varphi \tilde{\varphi} \varphi \tilde{\varphi} \rangle_{A, \parallel}^{(s)} \evaluated_{d=5} = & - \frac{2 (k_1 - k_2) (k_3 - k_4) }{s^2 E^3}
\eea
with the prefactor $(2\pi)^5 \delta^5 \big(\sum_{a=1}^4 \textbf{k}_a \big)$; and for the vectorlike scalar we have\footnote{
The functions $B_n$ in the transverse expression are defined as
\begin {align}
& B_3 = + 2 k_1 k_2 k_3 (k_1 + k_2 + k_3) (k_{12}-s) E_L E_R^2 \\
& B_2 = + \big[ 2k_1k_2 k_3 (s^2 +k_3 k_{12}) - (k_{12}-s) E_L k_{12} k_{13} k_{23} \big] E_R^2 \\
& B_1 = + \big[ k_{12}^4 - s^2 (k_1^2 + k_2^2 + k_3^2) + (k_{12}^2 + 2k_1 k_2) k_3^2 \big] E_R^2 \\
& B_0 = - \big[2k_1k_2 + E_L (k_{12}-s)\big] \big[(s+ k_3) sk_3 + (s^2 + k_3^2 ) E_R \big]. 
\end {align}
}
\bea[eq:sigmad5]
\langle \sigma \tilde{\sigma} \sigma \tilde{\sigma} \rangle_{A, \perp}^{(s)} \evaluated_{d=5} = &\frac{(\textbf{k}_1 - \textbf{k}_2) \cdot (\textbf{k}_3 - \textbf{k}_4) + s^{-2} (k_1^2 - k_2^2) (k_3^2 - k_4^2) }{(k_{12}-s)^2 E_L^2 E_R^2} 
\nn \\ &\x 
\bigg(\sum_{n=0}^3 \frac{B_n} {E^n} \bigg) \label{eq:sigmad5tran}
\\
\langle \sigma \tilde{\sigma} \sigma \tilde{\sigma} \rangle_{A, \parallel}^{(s)} \evaluated_{d=5} = & - \frac{ 2k_1 k_2 k_3 k_4 + (k_{12} k_3 k_4 + k_1 k_2k_{34} ) E + k_{12} k_{34} E^2 }{s^2 } 
\nn\\& 
\x \frac{(k_1 - k_2) (k_3 - k_4)}{E^3} \label{eq:sigmad5long}
\eea 
with the prefactor $(2\pi)^5 \delta^5 \big(\sum_{a=1}^4 \textbf{k}_a \big)$. As expected, the flat-space limit of both correlators give the same form of amplitude as in \eqref{eq:flatD4}, expressable in terms of six-dimensional Mandelstam variables (up to a pre-factor of $k_1 k_2 k_3 k_4$ for $\sigma$).

\subsection{Scalar-gravity}
In the previous subsection, we analyzed \emph{the conformally coupled} and the so-defined \emph{vectorlike} scalars, which are dual to boundary CFT operators with scaling dimensions $\frac{d+1}{2}$ and $d-1$, respectively. In this section, however, we will instead consider a new type of scalar that we will dub as ``tensorlike'', i.e.
\be
\text{tensorlike scalar: }\nu=\frac{d}{2}\;\rightarrow\;\De=d.
\ee
This class of scalars has trivial effective mass $\mu^2=0$, in which the minimally coupled scalar with $m=0$ is often referred to as the \textit{massless scalar} in literature. Besides the obvious observation that tensorlike scalars share the scaling dimension of the graviton with which they mix in the scalar-gravity interaction (making the resultant theory rather interesting), such scalars are also relevant as they may be numerically indistinguishable from the contributions of conserved vector fields in the boundary CFT.\footnote{More specifically, traditional numerical conformal bootstrap techniques cannot distinguish a ``fake'' scalar of scaling dimension of $d$ from a conserved vector with the scaling dimension $d-1$, as they induce indistinguishable contributions \cite{Karateev:2019pvw}.}

For the four-point correlation function of the external tensorlike scalars with graviton exchange, the leading-order perturbative contribution can be written down using the scalar-gravity interaction in \eqref{eq: graviton scalar interaction}. In the position space, the correlator reads as
\begin {multline}
\<\f (\textbf{x}_1) \f (\textbf{x}_2) \f (\textbf{x}_3) \f (\textbf{x}_4) \>_G = \int \frac{d z dz' d\textbf{x}^d d\textbf{x}'^d} {z^{d+1} z'^{d+1}} \\
\x \left(\bar{g}^{i_1 i_2} \bar{g}^{j_1j_2}\mathcal{T}_{i_1 j_1} (z, \textbf{x}; \textbf{x}_1, \textbf{x}_2)\right)\mathcal{G}_{i_2 j_2, l_2 m_2} (z, \textbf{x}; z', \textbf{x}') \\\x
\left(\bar{g}^{l_1 l_2} \bar{g}^{m_1 m_2} \mathcal{T}_{l_1 m_1} (z', \textbf{x}'; \textbf{x}_3, \textbf{x}_4)\right)
\end {multline}
where we define 
\begin {multline}
\mathcal{T}_{\mu\nu} (z, \textbf{x}; \textbf{x}_a, \textbf{x}_b) \coloneqq \partial_\mu \cG^\phi (z, \textbf{x}; \textbf{x}_a) \partial_\nu \cG^\phi (z, \textbf{x}; \textbf{x}_b)\\
- \frac{1}{2} \bar{g}_{\mu \nu} \bar{g}^{\alpha \beta} \partial_{\alpha} \cG^\phi (z, \textbf{x}; \textbf{x}_a) \partial_{\beta} \cG^\phi (z, \textbf{x}; \textbf{x}_b)
\end {multline}
for brevity.\footnote{
We remind the reader that the overall metric decomposes as $g_{\mu \nu} = g_{AdS} + h_{\mu \nu}$ with perturbative $h_{\mu \nu}$, and for clarity we use $\bar{g} = g_{AdS} $ in the expressions. Note also that we work in the axial gauge for the metric perturbation ($h_{zz} = 0, h_{zi} = 0$), and thus there is no $\bar{g}^{zz}$ contribution.} 

We proceed to decompose the correlator into its transverse and orthogonal components, as in \S~\ref{sec:orthdecom}. This analysis is conducted in momentum space, paralleling our prior approach. Our primary focus is on the $s$-channel, with the understanding that the $t$- and $u$-channels can be derived through suitable permutations. A modest amount of algebra yields the transverse and longitudinal parts of the four-point graviton-exchanging function as
\begin{subequations}
\label{eq: four point graviton-exchanging}
\begin{multline}
\langle \phi (\textbf{k}_1) \phi (\textbf{k}_2) \phi (\textbf{k}_3) \phi (\textbf{k}_4) \rangle_{G, \perp}^{(s)} \ = \textbf{k}_1^i \textbf{k}_2^j \textbf{k}_3^l \textbf{k}_4^m \\
\times \widehat{\mathcal{P}}_{ij, lm}^{(\perp,\perp)} (\bk_1 + \bk_2) \mathcal{M}_{\perp; \ 2}^{\nu_\phi, \nu_\phi, \nu_\phi, \nu_\phi; \nu_G} \label{eq:GtranF}
\end{multline}
and
\begin{align}
\langle \phi (\textbf{k}_1) \phi (\textbf{k}_2) \phi (\textbf{k}_3) \phi (\textbf{k}_4) \rangle_{G, \parallel}^{(s)} \ = \sum_{b=1}^2 \bar{\mathcal{M}}_{\parallel, b; \ 2}^{\nu_\phi, \nu_\phi, \nu_\phi, \nu_\phi; \nu_G}
\label{eq:GlongF}
\end{align}
upto an overall factor $(2\pi)^d \delta^d \left(\sum_{a=1}^{4} \textbf{k}_a \right)$.\footnote{
Remember that the projectors and the transverse seed amplitude are defined in \eqref{eq:Gpro} and \eqref{eq:Mtran} respectively.
} The \emph{modified} longitudinal seed amplitude above is defined as 
\end{subequations}
\begin{multline}
\hspace*{-1.6em}
\bar{\mathcal{M}}_{\parallel, b; \ 2}^{\nu_\phi, \nu_\phi, \nu_\phi, \nu_\phi; \nu_G} = 2 \textbf{k}_{1}^{i_1} \textbf{k}_{2}^{j_1} \textbf{k}_3^{l_1} \textbf{k}_4^{m_1} \widehat{\mathcal{P}}_{i_1j_1, l_1m_1}^{b} \int _0^\infty d p \mathcal{K}\mathcal{K} \mathcal{J}_{\nu_\phi, \nu_\phi, \nu_G}^{(2)} (k_1, k_2, p) 
\\\x
f^{b} \mathcal{K}\mathcal{K} \mathcal{J}_{\nu_\phi, \nu_\phi, \nu_G}^{(2)} (k_3, k_4, p) 
\\
+\textbf{k}_1^{i} \textbf{k}_2^{j} \widehat{\mathcal{P}}_{ij, ll}^{b} \int _0^\infty d p \mathcal{K} \mathcal{K} \mathcal{J}_{\nu_\phi, \nu_\phi, \nu_G}^{(2)} (k_1, k_2, p)
\\\x f^{b} \overline{\mathcal{K} \mathcal{K} \mathcal{J}}_{\nu_\phi, \nu_\phi, \nu_G}^{(2)} (k_3, k_4, p) 
\\
+ \textbf{k}_3^{l} \textbf{k}_4^{m} \widehat{\mathcal{P}}_{ii, lm}^{b} \int _0^\infty d p \overline{\mathcal{K} \mathcal{K} \mathcal{J}}_{\nu_\phi, \nu_\phi, \nu_G}^{(2)} (k_1, k_2, p)
\\\x
f^{b} \mathcal{K}\mathcal{K} \mathcal{J}_{\nu_\phi, \nu_\phi, \nu_G}^{(2)} (k_3, k_4, p)
\\+ \frac{1}{2} \widehat{\mathcal{P}}_{ii, ll}^{b} \int_0^\infty dp \overline{ \mathcal{K} \mathcal{K} \mathcal{J}}_{\nu_\phi, \nu_\phi, \nu_G}^{(2)} (k_1, k_2, p)
\\\x
f^{b} \overline{\mathcal{K} \mathcal{K} \mathcal{J}}_{\nu_\phi, \nu_\phi, \nu_G}^{(2)} (k_3, k_4, p),
\end{multline} 
where the $\overline{ \mathcal{K} \mathcal{K} \mathcal{J}}$ functions contain z-derivatives of the Bessel $K$ functions,\footnote{We define $\overline {\mathcal{K}\mathcal{K} \mathcal{J}}_{\nu_\phi, \nu_\phi, \nu_G}^{(n)} (\textbf{k}_a, \textbf{k}_b, p)$ as
\be 
{}&\overline {\mathcal{K}\mathcal{K} \mathcal{J}}_{\nu_\phi, \nu_\phi, \nu_G}^{(n)} (\textbf{k}_a, \textbf{k}_b, p) \coloneqq - \textbf{k}_a \cdot \textbf{k}_b \mathcal{K}\mathcal{K} \mathcal{J}_{\nu_\phi, \nu_\phi, \nu_G}^{(n)} (k_a, k_b, p) \\{}& + \int_0^\infty \frac{dz}{z^{d-3}} \partial_{z} \bigg[\sqrt{\frac{2}{\pi}} (k_a z)^{\frac{d}{2}} K_{\frac{d}{2}} (k_a z) \bigg] \partial_{z} \bigg[ \sqrt{\frac{2}{\pi}} (k_b z)^{\frac{d}{2}} K_{\frac{d}{2}} (k_b z) \bigg] \bigg[z^{\frac{d}{2}-2} J_{\frac{d}{2}} (p z) \bigg]
\ee 
for $\nu_\phi = \nu_G = \frac{d}{2}$.
} and where $\widehat{\mathcal{P}}$ and $f^b$ are abbreviations for brevity.\footnote{The projector sums here are defined as $\widehat{\mathcal{P}}_{ij, lm}^{b=1} = \widehat{\mathcal{P}}_{ij, lm}^{(\perp, \parallel)} + \widehat{\mathcal{P}}_{ij, lm}^{(\parallel, \perp)}+ \widehat{\mathcal{P}}_{ij, lm}^{(\parallel, \parallel)}$ and $\widehat{\mathcal{P}}_{ij, lm}^{b=2} = \widehat{\mathcal{P}}_{ij, lm}^{(\parallel, \parallel)}$; using definition \eqref{eq:Gpro}, we can compute the sums whose indices are partially or fully contracted
\begin {align}
& \widehat{\mathcal{P}}_{ij, ll}^{b=1} = - \frac{2}{d-1}\eta_{ij}, \quad \widehat{\mathcal{P}}_{ii, lm}^{b= 1} = - \frac{2}{d-1} \eta_{lm} \\
& \widehat{\mathcal{P}}_{ij, ll}^{b=2} (\textbf{k}) = + \frac{2 (d-2)}{(d-1)} \frac{\textbf{k}_i \textbf{k}_j}{k^2}, \quad \widehat{\mathcal{P}}_{ii, lm}^{b=2} (\textbf{k}) = + \frac{2 (d-2)}{(d-1)} \frac{\textbf{k}_l \textbf{k}_m}{k^2} \\
& \widehat{\mathcal{P}}_{ii, ll}^{b=1} = - \frac{2 d} {(d-1)}, \quad \widehat{\mathcal{P}}_{ii, ll}^{b=2} = + \frac{2 (d-2)}{(d-1)}.
\end {align}
In addition, the $f^{b}$ functions are defined as $f^{1} = \frac{1}{p}$ and $f^{2} =\frac{k^2} {p^3}$.} 

This direct perturbative approach of the computation of this four-point correlation function can be compared to many other existing techniques in literature. In particular, similar decomposition for $d=3$ is considered in \cite{Ghosh:2014kba}, which follows from the decomposition of the $3d$ action into the transverse and longitudinal parts \cite{Liu:1998ty}. Our final numerical evaluations of the correlator agree for $d=3$ and also share the same formal expression of the transverse part, and it would be interesting to explore if their decomposition also generalizes to arbitrary $d$ in the same way our decomposition does. Other techniques include: Arnowitt-Deser-Misner (ADM) and \textit{in-in} formalism \cite {Seery:2006vu, Seery:2008ax}, bootstrap formalism with weight-shifting operators \cite{Arkani-Hamed:2018kmz}, lifting formalism from flat space \cite{Bonifacio:2022vwa, Baumann:2021fxj}, bootless bootstrap formalism \cite{Bonifacio:2022vwa}\footnote{
At $d=3$, we identify our transverse and longitudinal parts with the following structures in the bootstrap computation
\begin {align}
& 12 \big(\textbf{k}_1^i \textbf{k}_2^j \textbf{k}_3^l \textbf{k}_4^m\big) \widehat{\mathcal{P}}_{ij, lm}^{(\perp, \perp)} \mathcal{M}_{\perp; 2}^{\nu_\phi, \nu_\phi, \nu_\phi, \nu_\phi; \nu_G} = s^4 \Pi_{2, 2}^{(s)} f_{2,2}^{(s)} \\
& 12 \sum_{b=1}^2 \bar{\mathcal{M}}_{\parallel, b; \ 2}^{\nu_\phi, \nu_\phi, \nu_\phi, \nu_\phi; \nu_G} = \big[s^2 \Pi_{2,1}^{(s)} f_{(2,1)}^{(s)} + (E_L E_R - s E) \Pi_{2,0}^{(s)} f_{(2,0)}^{(s)} + f_c \big] 
\end {align}
where the polarization tensors $\Pi_{2, i}$ as well as functions $f_{(2,i)}$ and $f_c$ are defined in \cite{Bonifacio:2022vwa}. In particular, we find $f_{2,2}^{(s)} =\mathcal{M}_{\perp; 2}^{\nu_\phi, \nu_\phi, \nu_\phi, \nu_\phi; \nu_G}$.}, and double-copy formalim with correction terms \cite{Armstrong:2023phb}. The final evaluation of the correlator $\langle \phi \phi \phi \phi \rangle_{G}$ from all these techniques, including our version and the previous version of orthogonal decomposition, have been numerically cross-checked and have been found to match for AdS$_4$.

Unlike AdS$_4$, there are not so many formalisms within which the scalar gravity correlation functions have been computed for AdS$_6$. To complement this, we provide the explicit evaluation of $\langle \phi \phi \phi \phi \rangle_{G}$ in AdS$_6$ (together with the evaluation in AdS$_4$) in a Mathematica file attached to this paper. We have explicitly checked that our correlators at $d=3, 5$ satisfy the manifestly local test (MLT) \cite{Jazayeri:2021fvk}; interestingly, our transverse and longitudinal components pass the MLT separately. 

\subsection{Pure Yang-Mills}

As an important example of using perturbation theory to compute external spinning correlators in momentum space, let us examine the tree-level four-point gluon correlator $\langle JJJJ \rangle$ in AdS$_{d+1}$. The full tree-level four-point gluon correlator $\langle JJJJ \rangle$ in AdS$_{d+1}$, associated with action \eqref{eq: pure YM and gravity}, consists of three pieces, the $s$-channel expression, the $t$-channel expression and a contact-diagram expression\footnote{Note that in a color-ordered correlator, there is no $u$-channel.}
\be 
\langle JJJJ \rangle = \langle JJJJ \rangle^{(s)}+ \langle JJJJ \rangle^{(t)}+ \langle JJJJ \rangle^{(c)} \ ,
\ee 
where the channels are exchanging a gluon in the bulk. In particular, the formal expression of the gluon-exchanging contribution here directly generalizes the scalar-gluon version \eqref{eq:scalargluon}, with the external bulk-to-boundary scalar propagators replaced by their gluon counterparts. As usual, we use the Poincar\'e coordinates and choose the axial gauge, and we will be suppressing the color indices.

The perturbative expression of correlator in the leading order can be expressed with the gluon bulk-to-boundary propagator \eqref{eq:gluonbtoB} and the gluon bulk-to-bulk propagator \eqref{eq:gluonbtob}. After some algebra, we find the following $s$-channel contribution in momentum space
\begin {multline}
\langle JJJJ \rangle^{(s)} = \int_0^\infty dp \mathcal{K} \mathcal{K} \mathcal{J}_{\nu_A, \nu_A, \nu_A}^{(1)} (k_1, k_2, p) \epsilon_i (\bk_1) \epsilon_j (\bk_2) \mathcal{V}^{ijm}_{\bk_1,\bk_2,-\bk} \\
\times \bigg(\frac{-i p H_{mn} (\textbf{k}, p)}{s^2 + p^2} \bigg) \\
\times \epsilon_k (\bk_3) \epsilon_l (\bk_4) \mathcal{V}^{kln}_{\bk_3,\bk_4,\bk} \mathcal{K}\mathcal{K} \mathcal{J}_{\nu_A, \nu_A, \nu_A}^{(1)} (k_3, k_4, p),
\end {multline}
where the $\mathcal{K} \mathcal{K} \mathcal{J}$ function is defined in \eqref{eq: definition of KKJ}, $\nu_A$ is $\frac{d}{2} -1$, and the momentum conservation gives $\bk= \bk_1 + \bk_2 = - \bk_3 - \bk_4$. In addition, the vertex operator here $\mathcal{V}_{\bk_1, \bk_2, \bk}^{ijm} = z_1^{-4} V_{\bk_1, \bk_2, \bk}^{ijm}$ is stripped off of z-powers (see full vertex definition in \eqref{eq:vertex factors for YM and Gravity}). As before, we express the correlator as a sum of transverse and longitudinal components
\begin {multline}
\langle JJJJ \rangle^{(s)} = (2\pi)^d \delta^d \bigg(\sum_{a=1}^4 \textbf{k}_a \bigg) \epsilon_i (\bk_1) \epsilon_j (\bk_2) \epsilon_k (\bk_3) \epsilon_l (\bk_4) \\
\mathcal{V}^{ijm}_{\bk_1,\bk_2, -\bk_1 - \bk_2} \mathcal{V}^{kln}_{\bk_3, \bk_4, \bk_1 + \bk_2} \bigg( \widehat{P}_{mn}^{\perp} \mathcal{M}_{\perp; \ 1}^{\nu_A, \nu_A, \nu_A, \nu_A; \nu_A} + \widehat{P}_{mn}^{\parallel} \mathcal{M}_{\parallel; \ 1}^{\nu_A, \nu_A, \nu_A, \nu_A; \nu_A}\bigg)
\end {multline}
where the projectors are defined in \eqref{eq:projectors}, and the transverse as well as the longitudinal seed amplitudes are defined in \eqref{eq:Mtran} and \eqref{eq:Mlong}, respectively. The $t$-channel expression can be extracted from the above result via the usual permutation 
\be 
\langle JJJJ \rangle^{(t)} = \langle JJJJ \rangle^{(s)}\evaluated_{1\to2\to3\to4\to1}.
\ee 
On the other hand, the contact diagram is simply given by
\begin{align}
\langle JJJJ \rangle^{(c)}=\int_0^\infty \frac{dz_1}{z_1^{d+1}} z_1^4 \mathcal{V}^{ijkl} \mathcal{G}_i \mathcal{G}_j \mathcal{G}_k \mathcal{G}_l, 
\end {align}
where $\mathcal{G}_{i}$ in this expression are the gluon bulk-to-boundary propagators \eqref{eq:gluonbtoB}, and $(z_1^4 \mathcal{V}^{ijkl}) = V^{ijkl}$ is the relevant vertex factor defined in \eqref{eq:vertex factors for YM and Gravity}.

We show two examples of evaluation as usual. For AdS$_4$, the transverse and the longitudinal seed amplitudes in the $s$-channel are
\be 
\mathcal{M}_{\perp; \ 1}^{\nu_A, \nu_A, \nu_A, \nu_A; \nu_A} \bigg|_{d=3} = \frac{1}{E_L E_R E}, \quad \mathcal{M}_{\parallel; \ 1}^{\nu_A, \nu_A, \nu_A, \nu_A; \nu_A} \bigg|_{d=3} = \frac{1}{k_{12} k_{34} E};
\ee 
and the contact-diagram expression becomes
\be 
\langle JJJJ \rangle^{(c)} \bigg|_{d=3} & = \frac{1}{E}.
\ee 
In fact, the above seed amplitudes are identical to those in \eqref{eq:scalarqedd3}, which are the seed amplitudes of $\langle \sigma \tilde{\sigma} \sigma \tilde{\sigma} \rangle_A^{(s)} = \langle \varphi \tilde{\varphi} \varphi \tilde{\varphi} \rangle_A^{(s)}$, because the vectorlike scalar has the same scaling dimension as the spin-1 gauge field. 

The AdS$_6$ seed amplitudes of $\langle JJJJ \rangle$ in the $s$-channel are
\begin {align}
\mathcal{M}_{\perp; \ 1}^{\nu_A, \nu_A, \nu_A, \nu_A; \nu_A} \bigg|_{d=5} & = \frac{1}{ (k_{12}-s)^2 E_L^2 E_R^2 } \bigg(\sum_{n=0}^3 \frac{A_n}{E^n} \bigg), \\
\mathcal{M}_{\parallel; \ 1}^{\nu_A, \nu_A, \nu_A, \nu_A; \nu_A} \bigg|_{d=5} & = \frac{2 k_1 k_2 k_3 k_4 + (k_{12} k_3 k_4 + k_1 k_2k_{34} ) E + k_{12} k_{34} E^2 }{k_{12} k_{34} E^3 } 
\end {align}
which are identical to the seed amplitudes of $\langle \sigma \tilde{\sigma} \sigma \tilde{\sigma} \rangle_A^{(s)}$ in \eqref{eq:sigmad5}; in comparison, the contact-diagram expression reads\footnote{The $g_n$ functions here are defined as 
\bea 
g_3 = & -2 k_1 k_2 k_3 (k_1 + k_2 +k_3) \\
g_2 = & - (k_1^2 + k_2 k_3) k_{23} + k_1 (k_2^2 + k_3^2) \\
g_1 = & - (k_1^2 + k_2^2 + k_3^2) \\
g_0 = & - k_4. 
\eea}
\be 
\langle JJJJ \rangle^{(c)} \bigg|_{d=5} = \sum_{n=0}^3 \frac{g_n} {E^n}.
\ee 
The equivalence of the seed amplitudes between $\langle JJJJ \rangle$ and $\langle \sigma \tilde{\sigma} \sigma \tilde{\sigma} \rangle_A$ reflects the value of studying vectorlike scalars when it comes to computing correlation functions for spin-1 gauge fields. A reasonable extension of such observation is the equivalence of the seed amplitudes between $\langle TTTT \rangle$, the graviton four-point function, and $\langle \phi \phi \phi \phi \rangle_G$ which we computed in \eqref{eq: four point graviton-exchanging}. It is then of considerable interest to directly generate spinning correlators from the associated scalar functions. Recent progress in this direction include \cite{Bonifacio:2022vwa, Lee:2022fgr, Li:2022tby}, which rely on the bootstrap techniques and/or the application of weight-shifting operators. From another perspective, one can also take advantage of the dimensional map we introduced in section \ref{sec:class} and observe that the two-derivative seed amplitudes of $\langle \phi \phi \phi \phi \rangle_\phi$ in AdS$_{d-1}$ are equivalent to those of $\langle JJJJ \rangle$ in AdS$_{d+1}$ and $\langle TTTT \rangle$ in AdS$_{d-1}$.\footnote{The scalar function here is the highest $z$-power contribution from \eqref{eq:gzz} and \eqref{eq:gii} associated with bulk Lagrangian of the form $\int (\nabla \phi)^2 \phi$. }

\section{Conclusion}
In this study, we have concentrated on momentum space AdS correlators across multiple dimensions, with a specific focus on QED, Yang-Mills, and exchange graviton correlators in (Anti-)de Sitter space. We have provided explicit examples of correlators for scalar QED in both three and five boundary dimensions. Furthermore, our venture into five-dimensional analysis reveals concrete cases of exchange graviton correlators, which are meticulously detailed in the accompanying Mathematica files. Despite their complexity, these expressions align seamlessly with the expected flat space limits, thus underscoring the robustness of our approach. Additionally, our observations on the interrelation of scalar factors across different dimensions potentially offer robust computational and organizational tools for future studies in the field. 

In terms of the interrelation of theories, our explicit computations across various theories in different dimensions are set to also expand the application scope of the double copy concept within curved spacetime \cite{Albayrak:2020fyp, Armstrong:2020woi, Farrow:2018yni, Armstrong:2023phb, Diwakar:2021juk, Zhou:2021gnu, Jain:2021qcl, Alday:2022lkk, Herderschee:2022ntr, Cheung:2022pdk, Lee:2022fgr}. The double copy methodology proves to be instrumental in determining amplitudes in complex theories by utilising simpler theories as foundational inputs, and we hope that our results can be leveraged as valuable input for this research.

The extension and application of loop-level AdS computations represent a natural progression, as underscored by recent advances in the field \cite{Giombi:2017hpr, Albayrak:2020bso, Alday:2021ajh, Heckelbacher:2022fbx, Herderschee:2021jbi, Stawinski:2023qtw}. Furthermore, there is a compelling prospect in broadening our classification to include a more diverse range of theories. This expansion is particularly pertinent to the cosmological bootstrap initiative and the computation of de Sitter space correlators, as referenced in several recent studies \cite{Arkani-Hamed:2018kmz,Baumann:2019oyu,Baumann:2020dch,Sleight:2019hfp, Sleight:2021plv,Albayrak:2023hie}. Efforts in this direction are currently in progress.

\begin{acknowledgments}
We thank Jinwei Chu and Hayden Lee for conversations.
\end{acknowledgments}


\bibliography{collectiveReferenceLibrary,References} 

\providecommand{\href}[2]{#2}\begingroup\raggedright\begin{thebibliography}{10}

\bibitem{Elvang:2013cua}
H.~Elvang and Y.-t. Huang, \href{http://arxiv.org/abs/1308.1697}{{\ttfamily
  arXiv:1308.1697 [hep-th]}}.

\bibitem{Buonanno:2022pgc}
A.~Buonanno, M.~Khalil, D.~O'Connell, R.~Roiban, M.~P. Solon, and M.~Zeng,,
  ``{Snowmass White Paper: Gravitational Waves and Scattering Amplitudes},'' in
  {\em {Snowmass 2021}}.
\newblock 4, 2022.
\newblock \href{http://arxiv.org/abs/2204.05194}{{\ttfamily arXiv:2204.05194
  [hep-th]}}.

\bibitem{Arkani-Hamed:2010zjl}
N.~Arkani-Hamed, J.~L. Bourjaily, F.~Cachazo, S.~Caron-Huot, and J.~Trnka,
  \href{http://dx.doi.org/10.1007/JHEP01(2011)041}{{\em JHEP} {\bfseries 01}
  (2011) 041}.

\bibitem{Bern:2010ue}
Z.~Bern, J.~J.~M. Carrasco, and H.~Johansson,
  \href{http://dx.doi.org/10.1103/PhysRevLett.105.061602}{{\em Phys. Rev.
  Lett.} {\bfseries 105} (2010) 061602}.

\bibitem{Arkani-Hamed:2017tmz}
N.~Arkani-Hamed, Y.~Bai, and T.~Lam,
  \href{http://dx.doi.org/10.1007/JHEP11(2017)039}{{\em JHEP} {\bfseries 11}
  (2017) 039}.

\bibitem{Maldacena:1997re}
J.~M. Maldacena, \href{http://dx.doi.org/10.4310/ATMP.1998.v2.n2.a1}{{\em Adv.
  Theor. Math. Phys.} {\bfseries 2} (1998) 231--252}.

\bibitem{Witten:1998qj}
E.~Witten, \href{http://dx.doi.org/10.4310/ATMP.1998.v2.n2.a2}{{\em Adv. Theor.
  Math. Phys.} {\bfseries 2} (1998) 253--291}.

\bibitem{Albayrak:2018tam}
S.~Albayrak and S.~Kharel,
  \href{http://dx.doi.org/10.1007/JHEP02(2019)040}{{\em JHEP} {\bfseries 02}
  (2019) 040}.

\bibitem{Albayrak:2019yve}
S.~Albayrak and S.~Kharel,
  \href{http://dx.doi.org/10.1007/JHEP12(2019)135}{{\em JHEP} {\bfseries 12}
  (2019) 135}.

\bibitem{Albayrak:2019asr}
S.~Albayrak, C.~Chowdhury, and S.~Kharel,
  \href{http://dx.doi.org/10.1007/JHEP10(2019)274}{{\em JHEP} {\bfseries 10}
  (2019) 274}.

\bibitem{Albayrak:2020isk}
S.~Albayrak, C.~Chowdhury, and S.~Kharel,
  \href{http://dx.doi.org/10.1103/PhysRevD.101.124043}{{\em Phys. Rev. D}
  {\bfseries 101} no.~12, (2020) 124043}.

\bibitem{Albayrak:2020bso}
S.~Albayrak and S.~Kharel,
  \href{http://dx.doi.org/10.1103/PhysRevD.103.026004}{{\em Phys. Rev. D}
  {\bfseries 103} no.~2, (2021) 026004}.

\bibitem{Albayrak:2020fyp}
S.~Albayrak, S.~Kharel, and D.~Meltzer,
  \href{http://dx.doi.org/10.1007/JHEP03(2021)249}{{\em JHEP} {\bfseries 03}
  (2021) 249}.

\bibitem{Albayrak:2023jzl}
S.~Albayrak and S.~Kharel,
  \href{http://dx.doi.org/10.1007/JHEP05(2023)151}{{\em JHEP} {\bfseries 05}
  (2023) 151}.

\bibitem{Bzowski:2019kwd}
A.~Bzowski, P.~McFadden, and K.~Skenderis,
  \href{http://dx.doi.org/10.1103/PhysRevLett.124.131602}{{\em Phys. Rev.
  Lett.} {\bfseries 124} no.~13, (2020) 131602}.

\bibitem{Bzowski:2020kfw}
A.~Bzowski, P.~McFadden, and K.~Skenderis,
  \href{http://dx.doi.org/10.1007/JHEP01(2021)192}{{\em JHEP} {\bfseries 01}
  (2021) 192}.

\bibitem{Bzowski:2013sza}
A.~Bzowski, P.~McFadden, and K.~Skenderis,
  \href{http://dx.doi.org/10.1007/JHEP03(2014)111}{{\em JHEP} {\bfseries 03}
  (2014) 111}.

\bibitem{Bzowski:2022rlz}
A.~Bzowski, P.~McFadden, and K.~Skenderis,
  \href{http://dx.doi.org/10.1007/JHEP12(2022)039}{{\em JHEP} {\bfseries 12}
  (2022) 039}.

\bibitem{Marotta:2022jrp}
R.~Marotta, K.~Skenderis, and M.~Verma,
  \href{http://dx.doi.org/10.1007/JHEP03(2023)196}{{\em JHEP} {\bfseries 03}
  (2023) 196}.

\bibitem{Isono:2018rrb}
H.~Isono, T.~Noumi, and G.~Shiu,
  \href{http://dx.doi.org/10.1007/JHEP07(2018)136}{{\em JHEP} {\bfseries 07}
  (2018) 136}.

\bibitem{Isono:2019wex}
H.~Isono, T.~Noumi, and G.~Shiu,
  \href{http://dx.doi.org/10.1007/JHEP10(2019)183}{{\em JHEP} {\bfseries 10}
  (2019) 183}.

\bibitem{Coriano:2019nkw}
C.~Corian\`o, M.~M. Maglio, and D.~Theofilopoulos,
  \href{http://dx.doi.org/10.1140/epjc/s10052-020-8089-1}{{\em Eur. Phys. J. C}
  {\bfseries 80} no.~6, (2020) 540}.

\bibitem{Anand:2019lkt}
N.~Anand, Z.~U. Khandker, and M.~T. Walters,
  \href{http://dx.doi.org/10.1007/JHEP10(2020)095}{{\em JHEP} {\bfseries 10}
  (2020) 095}.

\bibitem{Jain:2021wyn}
S.~Jain, R.~R. John, A.~Mehta, A.~A. Nizami, and A.~Suresh,
  \href{http://dx.doi.org/10.1007/JHEP08(2021)089}{{\em JHEP} {\bfseries 08}
  (2021) 089}.

\bibitem{Jain:2021vrv}
S.~Jain, R.~R. John, A.~Mehta, A.~A. Nizami, and A.~Suresh,
  \href{http://dx.doi.org/10.1007/JHEP09(2021)041}{{\em JHEP} {\bfseries 09}
  (2021) 041}.

\bibitem{Armstrong:2023phb}
C.~Armstrong, H.~Goodhew, A.~Lipstein, and J.~Mei,
  \href{http://dx.doi.org/10.1007/JHEP08(2023)206}{{\em JHEP} {\bfseries 08}
  (2023) 206}.

\bibitem{Farrow:2018yni}
J.~A. Farrow, A.~E. Lipstein, and P.~McFadden,
  \href{http://dx.doi.org/10.1007/JHEP02(2019)130}{{\em JHEP} {\bfseries 02}
  (2019) 130}.

\bibitem{Coriano:2021nvn}
C.~Corian\`o, M.~M. Maglio, and D.~Theofilopoulos,
  \href{http://dx.doi.org/10.1140/epjc/s10052-021-09523-9}{{\em Eur. Phys. J.
  C} {\bfseries 81} no.~8, (2021) 740}.

\bibitem{Coriano:2020ees}
C.~Corian\`o and M.~M. Maglio,
  \href{http://dx.doi.org/10.1016/j.physrep.2021.11.005}{{\em Phys. Rept.}
  {\bfseries 952} (2022) 1--95}.

\bibitem{Mack:2009mi}
G.~Mack, \href{http://arxiv.org/abs/0907.2407}{{\ttfamily arXiv:0907.2407
  [hep-th]}}.

\bibitem{Penedones:2010ue}
J.~Penedones, \href{http://dx.doi.org/10.1007/JHEP03(2011)025}{{\em JHEP}
  {\bfseries 03} (2011) 025}.

\bibitem{Rastelli:2016nze}
L.~Rastelli and X.~Zhou,
  \href{http://dx.doi.org/10.1103/PhysRevLett.118.091602}{{\em Phys. Rev.
  Lett.} {\bfseries 118} no.~9, (2017) 091602}.

\bibitem{Rastelli:2017udc}
L.~Rastelli and X.~Zhou, \href{http://dx.doi.org/10.1007/JHEP04(2018)014}{{\em
  JHEP} {\bfseries 04} (2018) 014}.

\bibitem{Alday:2020lbp}
L.~F. Alday and X.~Zhou,
  \href{http://dx.doi.org/10.1103/PhysRevLett.125.131604}{{\em Phys. Rev.
  Lett.} {\bfseries 125} no.~13, (2020) 131604}.

\bibitem{Alday:2020dtb}
L.~F. Alday and X.~Zhou,
  \href{http://dx.doi.org/10.1103/PhysRevX.11.011056}{{\em Phys. Rev. X}
  {\bfseries 11} no.~1, (2021) 011056}.

\bibitem{Alday:2022lkk}
L.~F. Alday, V.~Gon\c{c}alves, and X.~Zhou,
  \href{http://dx.doi.org/10.1103/PhysRevLett.128.161601}{{\em Phys. Rev.
  Lett.} {\bfseries 128} no.~16, (2022) 161601}.

\bibitem{Alday:2021odx}
L.~F. Alday, C.~Behan, P.~Ferrero, and X.~Zhou,
  \href{http://dx.doi.org/10.1007/JHEP06(2021)020}{{\em JHEP} {\bfseries 06}
  (2021) 020}.

\bibitem{Kharel:2013mka}
S.~Kharel and G.~Siopsis, \href{http://dx.doi.org/10.1007/JHEP11(2013)159}{{\em
  JHEP} {\bfseries 11} (2013) 159}.

\bibitem{Costa:2014kfa}
M.~S. Costa, V.~Gon\c{c}alves, and J.~a. Penedones,
  \href{http://dx.doi.org/10.1007/JHEP09(2014)064}{{\em JHEP} {\bfseries 09}
  (2014) 064}.

\bibitem{Sleight:2017fpc}
C.~Sleight and M.~Taronna,
  \href{http://dx.doi.org/10.1007/JHEP06(2017)100}{{\em JHEP} {\bfseries 06}
  (2017) 100}.

\bibitem{Goncalves:2019znr}
V.~Gon\c{c}alves, R.~Pereira, and X.~Zhou,
  \href{http://dx.doi.org/10.1007/JHEP10(2019)247}{{\em JHEP} {\bfseries 10}
  (2019) 247}.

\bibitem{Bissi:2022mrs}
A.~Bissi, A.~Sinha, and X.~Zhou,
  \href{http://dx.doi.org/10.1016/j.physrep.2022.09.004}{{\em Phys. Rept.}
  {\bfseries 991} (2022) 1--89}.

\bibitem{Li:2023azu}
Y.-Z. Li and J.~Mei, \href{http://dx.doi.org/10.1007/JHEP07(2023)156}{{\em
  JHEP} {\bfseries 07} (2023) 156}.

\bibitem{Alday:2023kfm}
L.~F. Alday, V.~Gon\c{c}alves, M.~Nocchi, and X.~Zhou,
  \href{http://arxiv.org/abs/2307.06884}{{\ttfamily arXiv:2307.06884
  [hep-th]}}.

\bibitem{Chu:2023pea}
J.~Chu and S.~Kharel, \href{http://arxiv.org/abs/2311.06342}{{\ttfamily
  arXiv:2311.06342 [hep-th]}}.

\bibitem{Cheung:2017ems}
C.~Cheung, C.-H. Shen, and C.~Wen,
  \href{http://dx.doi.org/10.1007/JHEP02(2018)095}{{\em JHEP} {\bfseries 02}
  (2018) 095}.

\bibitem{Raju:2012zr}
S.~Raju, \href{http://dx.doi.org/10.1103/PhysRevD.85.126009}{{\em Phys. Rev. D}
  {\bfseries 85} (2012) 126009}.

\bibitem{Raju:2011mp}
S.~Raju, \href{http://dx.doi.org/10.1103/PhysRevD.83.126002}{{\em Phys. Rev. D}
  {\bfseries 83} (2011) 126002}.

\bibitem{Bzowski:2015pba}
A.~Bzowski, P.~McFadden, and K.~Skenderis,
  \href{http://dx.doi.org/10.1007/JHEP03(2016)066}{{\em JHEP} {\bfseries 03}
  (2016) 066}.

\bibitem{Liu:2018jhs}
J.~Liu, E.~Perlmutter, V.~Rosenhaus, and D.~Simmons-Duffin,
  \href{http://dx.doi.org/10.1007/JHEP03(2019)052}{{\em JHEP} {\bfseries 03}
  (2019) 052}.

\bibitem{Giombi:2016ejx}
S.~Giombi,, \href{http://dx.doi.org/10.1142/9789813149441_0003}{``{Higher Spin
  \textemdash{} CFT Duality},''} in {\em {Theoretical Advanced Study Institute
  in Elementary Particle Physics}: {New Frontiers in Fields and Strings}},
  pp.~137--214.
\newblock 2017.
\newblock \href{http://arxiv.org/abs/1607.02967}{{\ttfamily arXiv:1607.02967
  [hep-th]}}.

\bibitem{Baumann:2019oyu}
D.~Baumann, C.~Duaso~Pueyo, A.~Joyce, H.~Lee, and G.~L. Pimentel,
  \href{http://dx.doi.org/10.1007/JHEP12(2020)204}{{\em JHEP} {\bfseries 12}
  (2020) 204}.

\bibitem{Maldacena:2011nz}
J.~M. Maldacena and G.~L. Pimentel,
  \href{http://dx.doi.org/10.1007/JHEP09(2011)045}{{\em JHEP} {\bfseries 09}
  (2011) 045}.

\bibitem{Karateev:2019pvw}
D.~Karateev, P.~Kravchuk, M.~Serone, and A.~Vichi,
  \href{http://dx.doi.org/10.1007/JHEP06(2019)088}{{\em JHEP} {\bfseries 06}
  (2019) 088}.

\bibitem{Ghosh:2014kba}
A.~Ghosh, N.~Kundu, S.~Raju, and S.~P. Trivedi,
  \href{http://dx.doi.org/10.1007/JHEP07(2014)011}{{\em JHEP} {\bfseries 07}
  (2014) 011}.

\bibitem{Liu:1998ty}
H.~Liu and A.~A. Tseytlin,
  \href{http://dx.doi.org/10.1103/PhysRevD.59.086002}{{\em Phys. Rev. D}
  {\bfseries 59} (1999) 086002}.

\bibitem{Seery:2006vu}
D.~Seery, J.~E. Lidsey, and M.~S. Sloth,
  \href{http://dx.doi.org/10.1088/1475-7516/2007/01/027}{{\em JCAP} {\bfseries
  01} (2007) 027}.

\bibitem{Seery:2008ax}
D.~Seery, M.~S. Sloth, and F.~Vernizzi,
  \href{http://dx.doi.org/10.1088/1475-7516/2009/03/018}{{\em JCAP} {\bfseries
  03} (2009) 018}.

\bibitem{Arkani-Hamed:2018kmz}
N.~Arkani-Hamed, D.~Baumann, H.~Lee, and G.~L. Pimentel,
  \href{http://dx.doi.org/10.1007/JHEP04(2020)105}{{\em JHEP} {\bfseries 04}
  (2020) 105}.

\bibitem{Bonifacio:2022vwa}
J.~Bonifacio, H.~Goodhew, A.~Joyce, E.~Pajer, and D.~Stefanyszyn,
  \href{http://dx.doi.org/10.1007/JHEP06(2023)212}{{\em JHEP} {\bfseries 06}
  (2023) 212}.

\bibitem{Baumann:2021fxj}
D.~Baumann, W.-M. Chen, C.~Duaso~Pueyo, A.~Joyce, H.~Lee, and G.~L. Pimentel,
  \href{http://dx.doi.org/10.1007/JHEP09(2022)010}{{\em JHEP} {\bfseries 09}
  (2022) 010}.

\bibitem{Jazayeri:2021fvk}
S.~Jazayeri, E.~Pajer, and D.~Stefanyszyn,
  \href{http://dx.doi.org/10.1007/JHEP10(2021)065}{{\em JHEP} {\bfseries 10}
  (2021) 065}.

\bibitem{Lee:2022fgr}
H.~Lee and X.~Wang, \href{http://dx.doi.org/10.1103/PhysRevD.108.L061702}{{\em
  Phys. Rev. D} {\bfseries 108} no.~6, (2023) L061702}.

\bibitem{Li:2022tby}
Y.-Z. Li, \href{http://dx.doi.org/10.1103/PhysRevD.107.125018}{{\em Phys. Rev.
  D} {\bfseries 107} no.~12, (2023) 125018}.

\bibitem{Armstrong:2020woi}
C.~Armstrong, A.~E. Lipstein, and J.~Mei,
  \href{http://dx.doi.org/10.1007/JHEP02(2021)194}{{\em JHEP} {\bfseries 02}
  (2021) 194}.

\bibitem{Diwakar:2021juk}
P.~Diwakar, A.~Herderschee, R.~Roiban, and F.~Teng,
  \href{http://dx.doi.org/10.1007/JHEP10(2021)141}{{\em JHEP} {\bfseries 10}
  (2021) 141}.

\bibitem{Zhou:2021gnu}
X.~Zhou, \href{http://dx.doi.org/10.1103/PhysRevLett.127.141601}{{\em Phys.
  Rev. Lett.} {\bfseries 127} no.~14, (2021) 141601}.

\bibitem{Jain:2021qcl}
S.~Jain, R.~R. John, A.~Mehta, A.~A. Nizami, and A.~Suresh,
  \href{http://dx.doi.org/10.1007/JHEP07(2021)033}{{\em JHEP} {\bfseries 07}
  (2021) 033}.

\bibitem{Herderschee:2022ntr}
A.~Herderschee, R.~Roiban, and F.~Teng,
  \href{http://dx.doi.org/10.1007/JHEP05(2022)026}{{\em JHEP} {\bfseries 05}
  (2022) 026}.

\bibitem{Cheung:2022pdk}
C.~Cheung, J.~Parra-Martinez, and A.~Sivaramakrishnan,
  \href{http://dx.doi.org/10.1007/JHEP05(2022)027}{{\em JHEP} {\bfseries 05}
  (2022) 027}.

\bibitem{Giombi:2017hpr}
S.~Giombi, C.~Sleight, and M.~Taronna,
  \href{http://dx.doi.org/10.1007/JHEP06(2018)030}{{\em JHEP} {\bfseries 06}
  (2018) 030}.

\bibitem{Alday:2021ajh}
L.~F. Alday, A.~Bissi, and X.~Zhou,
  \href{http://dx.doi.org/10.1007/JHEP02(2022)105}{{\em JHEP} {\bfseries 02}
  (2022) 105}.

\bibitem{Heckelbacher:2022fbx}
T.~Heckelbacher, I.~Sachs, E.~Skvortsov, and P.~Vanhove,
  \href{http://dx.doi.org/10.1007/JHEP08(2022)052}{{\em JHEP} {\bfseries 08}
  (2022) 052}.

\bibitem{Herderschee:2021jbi}
A.~Herderschee, \href{http://arxiv.org/abs/2112.08226}{{\ttfamily
  arXiv:2112.08226 [hep-th]}}.

\bibitem{Stawinski:2023qtw}
S.~F. Stawinski, \href{http://arxiv.org/abs/2309.15059}{{\ttfamily
  arXiv:2309.15059 [hep-th]}}.

\bibitem{Baumann:2020dch}
D.~Baumann, C.~Duaso~Pueyo, A.~Joyce, H.~Lee, and G.~L. Pimentel,
  \href{http://dx.doi.org/10.21468/SciPostPhys.11.3.071}{{\em SciPost Phys.}
  {\bfseries 11} (2021) 071}.

\bibitem{Sleight:2019hfp}
C.~Sleight and M.~Taronna,
  \href{http://dx.doi.org/10.1007/JHEP02(2020)098}{{\em JHEP} {\bfseries 02}
  (2020) 098}.

\bibitem{Sleight:2021plv}
C.~Sleight and M.~Taronna,
  \href{http://dx.doi.org/10.1007/JHEP12(2021)074}{{\em JHEP} {\bfseries 12}
  (2021) 074}.

\bibitem{Albayrak:2023hie}
S.~Albayrak, P.~Benincasa, and C.~Duaso~Pueyo,
  \href{http://arxiv.org/abs/2305.19686}{{\ttfamily arXiv:2305.19686
  [hep-th]}}.

\end{thebibliography}\endgroup
\bibliographystyle{utphysModified}
\end{document}